\documentclass[preprint,12pt]{elsarticle}
\usepackage[pdftex]{hyperref}
\usepackage{graphicx}

\usepackage{amssymb}
\usepackage{geometry}
\usepackage[fleqn]{amsmath}
\usepackage{bm}
\usepackage{amsbsy}
\usepackage{lineno}
\usepackage{physics}
\usepackage{mathtools}
\usepackage{xparse}
\usepackage{accents}

\usepackage{natbib}
\newcommand{\imagi}{
\mathrm{i}
}

\newcommand{\classoperator}[2]{
\left[
\begin{array}{c}
#1\\ \\
#2
\end{array} \right]
}

\begin{document}
%\preprint{}

\title{Revisiting the Lagrange theory for isolated n-particle systems with variable masses connected by an unknown field}


\author{Israel A. Gonz\'alez Medina}
%\thanks{}
%For APS jornals
%\affiliation{Physics Department\\ Instituto Superior de Tecnologias y Ciencias Aplicadas.\\ Universidad de la Habana. \\Cuba}

%For elsevier
\address{Instituto Superior de Tecnologias y Ciencias Aplicadas.\\ Universidad de la Habana. \\Cuba}
\ead{israelariel.gonzalezmedina@gmail.com}


\date{\today}

\begin{abstract}
We propose a new classical approach for describing a system composed of $n$ interacting particles with variable mass connected by a single field with no predefined form ($n$-VMVF systems). Instead of assuming any particular nature or analytical function for representing the variation of the masses or field, we propose them as unknown functions dependent on the particle positions and velocities. The work presents the Lagrangian theory which incorporates such variations which are find using only first principles. 

The consideration of mass as unknown quantity lead us to modify the D'Alembert's principle to ensure the compliance of the relativity principle. Also, because the addition of new variables to the system, we add a new and independent set of Lagrange equations depending on the $3$-D angular coordinates for the system of equations remain solvable. The four-dimensional space-time naturally appears in the problem when the position of the particle is expressed as a function of angular coordinates. This transformation set the $3-$D space of the angular coordinates as the stereographic projection of the $4-$D sphere defined by the Lorentz condition in the space-time. 

We identify two sets of constraints, each one for every coordinate system, by forcing the system to satisfy the laws of the conservation of the linear and angular momentum. The $\ddot{x}$ dependency's of the constraints functions set the necessity of extending the classical theory up to the second order of the Lagrange function. The obtained constraints are added to the initial Lagrangian by the Lagrange multiplier method and obtain not one, but two Lagrange functions and with them, two set of Lagrange equations for finding the final solution.
\end{abstract}

%\keywords{}

\maketitle

\section*{Introduction}
The quantum Revolution took place in the first quarter of the twentieth century in the understanding of microscopic phenomena. Quantum mechanics not only replaced classical mechanics as the theory to explain them, but it also revised fundamental concepts of what most people know as reality, what some authors call ``the quantum mechanical way of thinking''. The quantum theory works fine until it deals with a relativistic particle, starting with photons which have rest mass zero, and correspondingly travel in the vacuum at speed $c$. 

Once the conceptual framework of quantum mechanics was established, theoreticians direct their efforts to extend quantum methods to fields and particles. The most successful outcome of these works was the inception of quantum field theory (QFT) which provide a suitable procedure for quantizing fields and defining particles as excited states of such fields. QFT evolves and together with other theories like the electroweak theory and chromodynamics conforms what is known today as the Standard Model (SM). SM and also different approaches include spontaneous symmetry breaking whose distinctive particle, the Higgs particle was finally detected in 2012 at CERN. The SM successfully classifies all known elementary particles and describes all fundamental interactions except gravity. Besides the challenge of including gravitation to the theory, SM presents some deficiencies such as the strong CP problem, neutrino oscillations, matter-antimatter asymmetry, and the nature of dark matter and dark energy \citep{Ellis:1457821}.

Other theories have been proposed, not without including their own drawbacks and controversies, to solve the entirety of current phenomena such as the Minimal Supersymmetric Standard Model (MSSM) and Next-to-Minimal Supersymmetric Standard Model (NMSSM), and entirely novel explanations, such as string theory, M-theory, and extra dimensions. 

Despite the achievements and deficiencies of each model, the major problem is the lack of a single, all-encompassing and coherent theoretical framework that fully explain the interaction of particles with fields at high energies.

The success of the quantum mechanic theory is undisputed, pointing to a trustworthy strategy to follow when we study any physical phenomena at the quantum length scale.
The theoretical bases of QFT are indeed extracted from the quantum theory. However, modern quantum mechanics lies in classical mechanics as the theory from where the physical concepts are included. This fact can be troublesome since the Lagrange, and Hamilton's classical theory is developed considering constant masses while QFT intents to describe variable mass phenomena.

Recently, some authors have the opinion that rest mass should not be treated as a fixed quantity. An interesting review of this topic is found in the Journal of Physics by Mark Davidson \cite{Davidson2014, Davidson2015}. The author cites some unusual nuclear reactions in condensed matter that can be explained assuming the variation of the rest mass. An important conclusion extracted from these works is that variable mass theories are compatible with the requirement of general relativity.

It is our understanding that any quantum mechanic based theories like QFT, constructed to describe phenomena where masses are variables, should be built on a quantum theory based on a classical approach that considers the mass as a variable quantity.

On the other side, the fields and their couplings are included \textit{a priori} on most of the theories in physics, $e.i.$ their expressions are already predefined. The fundamental interactions in nature are identified by the relative strength of the force, the range of effectiveness, the kind of particles that feel the effect and the nature of the particles that mediate the force. However, these qualitative criteria are not enough for defining the expression of the interaction energy between particles needed for the Hamiltonian on the quantum theory. For example, even knowing the gluons as the carrier of the strong force, the hadrons as the particles that experience that force and the relative strength, the expression the strong force is unknown. Also, even knowing the expression for the force, its inclusion on the quantum theory can rise some difficulties, being the gravity the most notable example.

\subsection*{Main Tasks, methodology, and assumptions}

Our goal is to propose a quantum theory that set the fundamental basis for describing the phenomena of the interaction of particles with fields by including variables masses and removing the obstacle of \textit{a priori} potentials. Instead, the masses and the field should be found as a solution to the problem, using only first principles.

The construction of the modern Quantum Theory defines quantum states as vectors in an abstract complex vector space named the Hilbert space, whose dimension is determined by the degree of freedom of the physical system under study. The laws of physics are included from the canonical transformations of the classical Hamiltonian formulation, which at the same time are obtained from the Lagrangian theory. When a reader, who expect to include the mass as a variable quantity in the Quantum theory, follows the modern method for constructing such theory, questions like the following might come out: 
\begin{itemize}
\item Which is the quantum operator that changes the state described by the mass? Alternatively, what is the operator $\mathcal{M}$ whose action over such state kets is
\begin{equation}
\mathcal{M}(dm) | m_0 \rangle = | m_0 + dm\rangle ?
\end{equation}
also, if such an operator exists,
\item what would be its associated canonical transformation in the classical theory?
\end{itemize}

Those questions are the primary motivation for this work. Following the modern quantum mechanic approach, we define four tasks for the construction of the quantum theory for particles with variable mass:
\begin{enumerate}
\item develop a classical theory of Lagrange that includes the particle mass and field as unknown and variables quantities. 
\item develop a classical theory of Hamilton with the final goal of obtaining the infinitesimal canonical transformations of the variables of the system.
\item create a vector space, if needed, which satisfies the structure of the new classical theory.
\item construct a quantum theory using previous results, including and adapting new and existing quantum axioms following the ``the quantum mechanical way of thinking''. 
\end{enumerate}

This methodology agrees with Dirac at ``Lectures of Quantum mechanics'' \cite{dirac2001lectures} where he analyze the process of quantization and concludes by saying: ``I don't think one can in any way short-circuit the route of starting with an action integral, getting a Lagrangian, passing from the Lagrangian to the Hamiltonian, and then passing from the Hamiltonian to the quantum theory.'' 

The pre-print version of the entire proposal for the construction of the Quantum Theory including masses and field as unknown functions can be found in the article entitled ``\href{http://arxiv.org/abs/1811.12175}{A new proposal for a quantum theory for isolated n-particle systems with variable masses connected by a field with variable form}'' \cite{Israel:1811.12175}. This work covers the first item of the methodology of developing the Lagrange theory for such systems. 

\section{Revisiting the classical theory. \texorpdfstring{$n$}-VMVF systems.}\label{ClassicalLagSection}

The construction of a classical theory that considers the mass as a variable quantity must start from the very beginning. It means to start revisiting the Newton Laws, as the base of most of the classical theories. Variable masses with time, mostly depending on the velocity, has been studied on several problems of the classical mechanic, for example, the rocket problem. It is well-known that the mass variation of a single particle violates the relativity principle under a Galilean transformation. The Newton's second law states that the force $\mathbf{F}$ acting on a particle with mass $m$ and velocity $\mathbf{v} $ is equal to the time derivative of the momentum:
\begin{equation}
\mathbf{F} = \frac{d(m\mathbf{v})}{dt} = m\frac{d\mathbf{v}}{dt} + \mathbf{v} \frac{dm}{dt}. \label{Newtown2Law}
\end{equation} 
A contradiction appears in the presence of a null force $\mathbf{F}=\mathbf{0}$ if mass varies with time ($\frac{dm}{dt}\neq 0$). In effect, Newton's first law says that the particle remains at rest in a system where it is initially at rest, but, according to equation \ref{Newtown2Law}, it is also accelerated by a force of value  $-\dot{m}\mathbf{v}$ in a system where the particle moves with velocity $\mathbf{v}$ \cite{Plastino1992}. 

One way to solve this contradiction is to set constraints on the variation of mass with time. For example, by considering the mass lost as isotropic, so the last term of equation \ref{Newtown2Law} vanish. However, in this work, we propose to study the variation of the mass of the particle, and for that, we consider it as a degree of freedom, same as position. We can conclude then, based on the previous discussion that isolated particle whose mass varies with no restriction can no longer exist, and this is the first assumption of this approach. 

The relativity principle under a Galilean transformation is satisfied for the particle with variable mass if there is some external ``action'' to suppress the violation. On an isolated particle system with more than one particle, the external force acting over one particle must necessarily proceed from the other particles of the system. 

We can define then the field as the physical magnitude sourced from all particles that produce all the coordinated actions to preserve the zero net force of isolated particle systems. As the value field depends at any time on the position of the particles and its derivatives and the value and variation of the mass, then the field must also be considered as a variable and unknown quantity, to be found by the motion equations. We call those systems of $n$ particles with variable masses and connected by a variable field with no predetermined form as $n$-VMVF systems.

\begin{figure}[h!]
\centering
\includegraphics[width=0.3\textwidth]{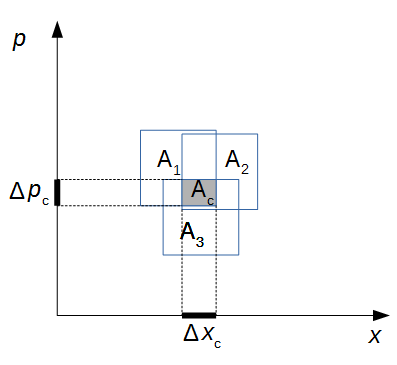}
\caption{Phase diagram for 3 particle at quantum scale.} \label{phaseDig}
\end{figure}

We assume that the response time of the system of those variations is zero or what is the same; we are proposing a system whose components interact via instantaneous interactions. We hypothesize that this behavior exists when length scale is in the order of the system's de Broglie wavelength $e.i$ at the quantum scale. In quantum mechanics, because of the Heisenberg's Uncertainty Principle, if the distance between particles is small enough, there should exist a spatial volume where the probability to find all the particle at the same point is different from zero. The phase diagram of figure \ref{phaseDig} 
represents the phase volume of tree particles at the quantum scale. Each particle corresponding to a phase volume $A_1$, $A_2$ and $A_3$ is displayed as equal squares of volume  $\hbar$. At close enough distance there will exist a phase volume shared by all particles, represented by the gray area $A_c$. On this subspace, each point has a not null probability of finding all particle together at the same time. We suppose that the existence of this commonplace volume allows the particle system to behaves as a single object, so its components properties, such as every particle position, masses, became intrinsic properties of a unique object. Under this point of view, masses and field derivatives will instantaneously vary in a ``harmonic way'', so the system can satisfy conservation laws as one single physical object.

The assumption of considering the mass of the particles and the field as unknown variables of the system cannot be traduced that they are generalized coordinates of the system or the coordinates that the Lagrange operator depends on. If they are treated as such, it implies that there are conservations laws that corresponds to the case when the Lagrangian do not depend on the masses or the field. For example, the absence of the rectangular or angular coordinates in the Lagrangian of an isolated system is related to the conservation of the linear and the angular momentum which reflects the space properties of homogeneity and isotropy, respectively. Having no knowledge of any similar conservation law related to an space property for the masses of the particle or the field, we state that the Lagrange function for $n$-VMVF systems depends only on the coordinates and its derivatives with time. Because of that, the mass of the particle and the field must be considered as function of the position of the particle and its derivatives with time like
\begin{align}
U \equiv& U(\mathbf{r}_1, \mathbf{r}_2, ..,\mathbf{\dot{r}}_1, \mathbf{\dot{r}}_2,...) 
\nonumber \\
m_n \equiv& m_n(\mathbf{r}_1, \mathbf{r}_2, ..,\mathbf{\dot{r}}_1, \mathbf{\dot{r}}_2,...).
\end{align}  

\section{The modification of D'Alembert's Principle for \texorpdfstring{$n$}-VMVF systems}

From the classic theory, the D'Alembert's principle states that \textit{``The total virtual work of the impressed forces plus the inertial forces vanishes for reversible displacements''}. The principle is written as a 
\begin{equation}
\sum_n (\mathbf{F}^{(a)}_n - \dot{\mathbf{p}}_n + \mathbf{f}_n)\cdot\delta \mathbf{r}_n = 0,
\end{equation}
where $\mathbf{F}^{(a)}_n$ are the applied forces, $\mathbf{f}_n$ are the constraint forces, and the dynamical effects are included by a ``reversed effective force'' $- \dot{\mathbf{p}}_n$. The method of transforming the dynamical problem into a static phenomenon by the inclusion of the ``reversed effective force'' is known in the literature as the Bernoulli and D'Alembert's method.

The restriction of set the net virtual work of the constraint forces to zero,
\begin{equation}
\sum_n \mathbf{f}_n\cdot\delta \mathbf{r}_n = 0 \qquad \text{or}
\qquad
\sum_n (\mathbf{F}^{(a)}_n - \dot{\mathbf{p}}_n)\cdot\delta \mathbf{r}_n = 0, \label{DAlembertOrig}
\end{equation}
is known in the literature as the D'Alembert's principle. The existence of constraint forces implies that $\delta \mathbf{r}_n$ are not entirely independent but connected by constraint equations. This fact means coefficients $\delta \mathbf{r}_n $  can be no longer zero in equation \ref{DAlembertOrig} $e.i$ $\mathbf{F}^{(a)}_n - \dot{\mathbf{p}}_n \neq 0$.

The D'Alembert's principle as stated, can no longer be applied to an isolated particle with variable masses because of the absence of constraints (independent $\delta \mathbf{r}_n $), lead to the compliance of the second Newton Law, which we saw cannot be applied on this problem. A particle system with variable masses, however, must satisfy that the sum of the net force acting on the particle, $\mathbf{F}_n$, must vanish as
\begin{equation}
\sum_n \mathbf{F}_n=0 \label{sysEquCond}.
\end{equation}

The force acting over every particle $\mathbf{F}_n$ can be divided into the applied and the constraint forces like
\begin{align}
&\mathbf{F}_n(\{\ddot{\mathbf{r}}_n\},\{\dot{\mathbf{r}}_n\}, \{\mathbf{r}_n)\} \equiv \mathbf{F}^{(a)}_n  - \dot{\mathbf{p}}_n + \mathbf{f}_n\neq 0.
\end{align}

We can divide the constrained forces into two types: the constrained forces related to geometric restrictions and the constraints related to the coordinated action between the particles of the system for the particle system with variable mass satisfy the zero net force condition. 

The first type of constraints groups those we are used to known, and they are usually related to geometrical restrictions. For example, those that: keeps the distance between particles constant in a rigid body, or maintain the gas molecules moving inside a container, or force particles to move on a particular surface. The constraints of the first type can be removed from the physical system, and they set relations between the geometric coordinates of the system generating the generalized coordinates. There are several classifications for this type of constraints. One of the most important, of such type of constraints are the holonomic constraints which relate the particle coordinates in the form
\begin{equation}
f(\mathbf{r}_1, \mathbf{r}_2,...\mathbf{r}_n, t)=0 \qquad \text{or} \qquad f(\{\mathbf{r}_n\}, t)=0, 
\end{equation}
and allow to express the position of the particles as
\begin{equation}
\mathbf{r}_n = \mathbf{r}_n(q_1,q_2...q_{3N-k},t),
\end{equation}
being $k$ the number of the constraint equations.

On the other side, the constraints related to the ``coordinate action'' between particles when masses vary taking into account are intrinsic of the system, and they can not be removed; otherwise, the system won't satisfy the zero net force condition. The constraint of the second type set relations to all the variables of the system, including masses and field derivatives. Such type of constraints relate all the variables of the system like
\begin{equation}
f(\{\mathbf{r}_n\}, \{\dot{\mathbf{r}}_n\}, \{m_n\}, \mathbf{A},t)=0.
\end{equation}
Note that is not possible to define independent generalized coordinates from intrinsic constraints such the particle's positions can be expressed in the form
\begin{equation*}
\mathbf{r}_n = \mathbf{r}_n(q_1,q_2...q_{3N-k},t).
\end{equation*}

We can separate then the constraints, $\mathbf{f}_n$, into the geometrical's $\mathbf{f}_n^{(g)}$ and the intrinsic constraints $\mathbf{f}_n^{(i)}$ like
\begin{equation}
\mathbf{f}_n = \mathbf{f}_n^{(g)} + \mathbf{f}_n^{(i)}.
\end{equation}
%On particle systems with no mass variation, a virtual particle displacement won't produce any virtual work over other particles positions, $\mathbf{F}_n \delta \mathbf{r}_{n'}$, even with an acting force depending on other particles positions because of the second Newton law relations its satisfied for each particle of the system   $\mathbf{F}_n=0$. In fact, it won't produce any virtual work event on its position.

If the particle system is in equilibrium; $e.i.$ the sum of all the forces acting on every particle vanishes, Eq. \ref{sysEquCond}, then the dot product $(\sum_n \mathbf{F}_n)\cdot \delta \mathbf{r}_{n'} $ also vanishes. This can be interpreted as the virtual work of the system given by the virtual displacement of any particle of the system in equilibrium is zero. Replacing the net forces with the applied forces, the geometrical and the intrinsic constraint forces, the virtual work of the system is
\begin{equation}
\sum_n (\mathbf{F}_n^{(a)} + \mathbf{f}_n^{(i)} - \dot{\mathbf{p}}_n)\cdot\delta \mathbf{r}_{n'} + \sum_n \mathbf{f}_n^{(g)}\cdot \delta \mathbf{r}_{n'} = 0.
\end{equation}

We propose then a modification of the D'Alembert Principle for particle systems where the second Newton law is not satisfied by the single particle. The modified principle states that: 
\begin{quote}
``The total virtual work of the sum of the impressed, the intrinsic constraint and the inertial forces vanishes for the reversible displacements of any particle of the system''.
\end{quote}

The new principle can be written as
\begin{equation}
\sum_n \mathbf{f}_n^{(g)}\cdot \delta \mathbf{r}_{n'} = 0 \label{DAlembertExt}.
\end{equation}
which led to the equation
\begin{equation}
\sum_n (\mathbf{F}_n^{(a)}  + \mathbf{f}_n^{(i)} - \dot{\mathbf{p}}_n )\cdot \delta \mathbf{r}_{n'} = 0.  \label{DAlembertExt1}
\end{equation}
The term constraint forces exclude those effects related to intern strengths that keep the zero net force condition of the system. Also, the principle does not apply for irreversible displacements, such as sliding friction.

Equation \ref{DAlembertExt1} can be separated according to the contributions from the constant and from the variable mass like
\begin{equation}
\sum_n \{\big(\mathbf{F}_n^{(a)} - \dot{\mathbf{p}}_n\big)_{[\dot{m}_n= 0]} 
+ \big(\mathbf{F}_n^{(a)}  + \mathbf{f}_n^{(i)} - \dot{\mathbf{p}}_n\big)_{[\dot{m}_n\neq 0]} \}\cdot \delta \mathbf{r}_{n'} = 0. \label{DAlembertExtSpl}
\end{equation}
The last term represents the dynamical, the intrinsic constraints and the applied force including mass variations while the first one includes only mass function dependency but not its derivatives. If the mass is constant for the first term, then
\begin{equation}
(\mathbf{F}_n^{(a)} - \dot{\mathbf{p}}_n\big)_{[\dot{m}_n= 0]}=0,
\end{equation}
then the solution can be found by solving the set of equations:
\begin{equation}
\sum_n \big(\mathbf{F}_n^{(a)}  + \mathbf{f}_n^{(i)} - \dot{\mathbf{p}}_n\big)_{[\dot{m}_n\neq 0]} \cdot \delta \mathbf{r}_{n'} = 0.
\end{equation}

From Lagrange's point of view, the D'Alembert's principle for $n$-VMVF systems of equation \ref{DAlembertExt1} means that the application of the Lagrange operator over the Lagrangian of the system, $L_{sys}$, using the degree of freedom $q$ of the particle $n$, results in
\begin{align}
\mathcal{L}_{q_{n}} L_{sys} &\equiv \Big[ \frac{d}{dt}\Big(\frac{\partial}{\partial \dot{q_n}} \Big) - \frac{\partial}{\partial q_n}\Big]L_{sys} \nonumber \\ 
&= \sum_{n'=1}^N \dot{P}_{q_{n'}} - Q_{q_{n'}}= 0 \;\;\;\; \forall\; n=1,2...,
\label{LagrangeForces}
\end{align}
where $\dot{P}_{q_{n'}}$, $ Q_{q_{n'}}$ are the generalized dynamical and applied force on particle $n'$, respectively, and where the last term also includes the intrinsic constraint since such constraints are also a type an applied force. For example, the equation of motion for an isolated 2-particle system in the  $x$-direction is:
\begin{align}
&\Big[ \frac{d}{dt}\Big(\frac{\partial}{\partial \dot{x}_{1}} \Big) - \frac{\partial}{\partial x_{1}}\Big]L_{sys}
= \Big[ \frac{d}{dt}\Big(\frac{\partial}{\partial \dot{x}_{2}} \Big) - \frac{\partial}{\partial x_{2}}\Big]L_{sys} 
\nonumber \\
&= \dot{P}_{x_{1}} + \dot{P}_{x_{2}} - Q_{x_{1}}  - Q_{x_{2}}=0 \label{MotionEqTrans}
\end{align}
The action of the Lagrangian operator depending on $x_1$ or $x_2$ over the Lagrange function $L$ results in the same expression, meaning that equation \ref{MotionEqTrans} is $n$-degenerated. This degeneration feature is the base idea to obtain the final Lagrangian function.

%These forces are the result of single particle Lagrangian operator acting over the Lagrangian of an isolated particle under the action of a potential considering mass variation, $L_{sp_{n'}}$,
%\begin{equation}
%\mathcal{L}_{q_{n'}} L_{sp_{n'}} = \dot{P}_{q_{n'}} - Q_{q_{n'}}.
%\end{equation}

We can divide the terms on those who contain the variation of the mass and those who do not, as shown equation \ref{DAlembertExtSpl}, and it suggests that the final Lagrangian can be split into two terms: one related to particle mass variations and other to the static masses:
\begin{equation}
L_{sys} = L_{[\dot{m}_n= 0]} + L_{[\dot{m}_n\neq 0]} = \sum_{n'=1}^N L_{sp_{n'}} + L_{[\dot{m}_n\neq 0]}. \label{splittedLagEq}
\end{equation}
The last term is the well known Lagrangian for interacting particles with constant masses system while $L_{[\dot{m}_n\neq 0]}$ vanishes for a constant value of the mass of the particle.

\section{Constructing Lagrangian for isolated n-VMVF systems. Initial assumptions.}

Let us start by defining the generalized coordinates of the system. For isolated $n$-VMVF systems there are no applied forces $\mathbf{F}_n^{(a)}$ and there are no geometric constraints $\mathbf{f}_n^{(g)}$ or equations connecting only coordinates. In that case, the Cartesian coordinates $x,y,z$ describe the system as the generalized coordinates. The D'Alembert principle of equation \ref{DAlembertExt1} take the form
\begin{equation}
\sum_n (\mathbf{f}_n^{(i)}({\{\dot{\mathbf{r}}_l}\}, \{\mathbf{r}_l\}) - \dot{\mathbf{p}}_n ({\{\dot{\mathbf{r}}_l}\}, \{\mathbf{r}_l\}))\cdot \delta \mathbf{r}_{n'} = 0, \label{DAlembertExt2}
\end{equation}
where $\mathbf{r}_l \equiv \mathbf{r}_l(x,y,z)$ and the generalized forces $Q_{q_{n'}}$ in equation \ref{LagrangeForces} includes only the intrinsic constraint forces.
 
As the next step, we can make use of the classification for the Lagrangian terms relative to the variation of mass as shown in equation \ref{splittedLagEq} and associate the term of constant mass, $ L_{[\dot{m}_n= 0]}$, with the last term in equation \ref{DAlembertExt2}. Form the ordinary classic mechanic; we know that the Lagrange function for an isolated particle system with constant mass that $ L = L_{[\dot{m}_n= 0]} = \sum_n \frac{1}{2}m_n\dot{\textbf{r}}^2_n$. The Lagrangian term in equation \ref{splittedLagEq} related to the variation of masses incorporates then only the intrinsic constraints as a unique field that connect the particles and ``transport'' the information between them. Let us represent this action with the potential energy $U$. In this case, the Lagrangian has the form:
\begin{equation}
L= \sum_n \frac{1}{2}m_n\	\dot{\textbf{r}}^2_n -U.
\end{equation}

From the extension of D'Alembert principle, we know that the result of the Lagrange equation is the sum of the applied, intrinsic and inertial forces over all particles or, the applied net force. We can find those forces if the $n$-particle system is approached to a $n$ systems of one particle.

Even we show that particles with variable mass cannot be isolated without violating universal laws, the system can be described by an observer at rest in an inertial frame by computing every particle motion. From the observer point of view, there is no way to say whether the particle is or isn't part of an isolated system. Only the observation of the phenomenon along the passing of time can tell if the set of particles behaves as the components of such a system. At any time, the inertial observer can describe the motion of every particle assuming that the particle is under the influence of the external field. This field depends on the position and velocities of the particle itself, but also of the other particles. At any time then, the observer can compute the net applied forces of every particle $\dot{P}_{n,i} -Q_{n,i}$ in any direction $i$. It is the zero summation of all measured forces during the evolution of the phenomenon that indicates the observer is in the presence of an isolated $n$-VMVF system.

Under the previous statement, a particle system can be considered as a set of one particle's systems where the forces acting on each one of them resume the action of the other particles. The functions of mass and the function of field energy in the one particle system have a dependency with variables like:
\begin{align}
U(\mathbf{r}_1, \mathbf{r}_2, ..,\mathbf{\dot{r}}_1, \mathbf{\dot{r}}_2,...) &\equiv U(\mathbf{r}_n, \alpha_1, ..,\mathbf{\dot{r}}_n, \beta_2,...)   
\nonumber \\
m_n(\mathbf{r}_1, \mathbf{r}_2, ..,\mathbf{\dot{r}}_1, \mathbf{\dot{r}}_2,...) &\equiv m_n(\mathbf{r}_n, \alpha_1, ..,\mathbf{\dot{r}}_n, \beta_2,...).
\end{align} 
being $\alpha_i$, $\beta_i$ the parameters related to the coordinates and velocities of the other particles.

The Lagrangian $L_{{sp}_n} $ for a single particle under the action of external forces derivable from the potential function $U$ is:
\begin{equation}
L_{{sp}_n} = \frac{1}{2}m_n\dot{\textbf{r}}^2_n -U.
\end{equation}

Consider a one particle system, the net applied and inertial forces obtained  from the Lagrange equations are
\begin{equation}
\dot{P}_{n,x} - Q_{n,x}  = \mathcal{L}_{n,x} L_{{sp}_n} 
 \equiv \Big[ \frac{d}{dt}\frac{\partial}{\partial \dot{x}_n} - \frac{\partial}{\partial x_n}\Big](\frac{1}{2}m_n\dot{\textbf{r}}^2_n -U ),
\end{equation}
where coordinates of particle $n$ are variables while the coordinates of the other particles are considered parameters. Because this fact, the generalized forces won't include the contribution of the derivatives of the field with those quantities. However, the  sum of the generalized forces over all particles indeed will include those contributions. Then, we can relate the approach of the $n$ one particle's system to the $n$-particle system by stating that  
\begin{quote}\label{genForceApproach}
``The sum of the net applied and generalized inertial forces of an isolated $n$ particle system is zero and equal to the sum of the net applied and inertial generalized forces acting on each particle if it is considered as a one particle system under the action of a field generated by the other particles.''
\end{quote}

In other words, the assumption means that
\begin{equation}
\sum_n \dot{P}_{n,x}^{(1)} - Q_{n,x}^{(1)} = \sum_n \dot{P}_{n,x}^{(N)} - Q_{n,x}^{(N)}, \label{ApproachAppliedForces}
\end{equation} 
where the super index $(1)$ indicate the net and inertial forces computed by the model of one particle system while the super index $(N)$ is referred to the expression obtained using the model of the system composed of $N$-particles.

\subsection{The functions for the mass and field}

The equations of motions must be solvable, which means the numbers of variables must match the number of independent equations. To accomplish this objective and also to reduce the complexity of the solution, we need to make some assumptions for the mass and field functions. A more specific analysis should be performed once we have defined the equations of motions of the system.

\begin{description}\label{classicalAssump1}
\item Mass

We consider the mass variation generated by internal changes in the structure of the particle depending only on the own particle's position, $e.i$
\begin{equation}
m_n \equiv m_n(\mathbf{r}_n).
\end{equation}
This type of variation of the internal variations is also shared by Davidson \citep{Davidson2014}, which presumes that the mass densities depends only on the particle position. It differs from the variations that appear at relativistic energies, which depends on the velocity of the particle. The derivatives of the function of mass then satisfy
\begin{align}
&\frac{\partial^{(l)} m_{n'}}{\partial x_{i_n}^{(l)}} \equiv
\frac{\partial^{(l)} m_{n'}}{\partial x_{i_n}^{(l)}} \delta_{nn'},
\;
\frac{\partial^{(l)} m_{n'}}{\partial \dot{x}_{i_n}^{(l)}}=0,
\end{align}
where $x_i = \{x,y,z\}$. 

\item Field

We proposed the field is the medium to connect particle, and nothing is said about its nature which we assume undefined. However, the undoubted success and abstract concepts introduced in the electromagnetic theory led us to include some of its basics ideas. 
We can presume the existence of two fields generated from all the particles of the system: one scalar $\phi$ and other vectorial $\mathbf{A}$. We propose the form of the potential energy for the field connecting particles in the system as
\begin{equation}
U_n = q\mathbf{A}\cdot \dot{\mathbf{r}}_n -q\phi,  \label{PotentialEnergyForm}
\end{equation}
where fields $\mathbf{A}$ and $\phi$  are considered to be time independent. In this case, the proposition for the Lagrange function for particle $n$ under the action of an external field, $L_{{sp}_n}$, have the form:
\begin{equation}
L_{{sp}_n} = \frac{1}{2}m_n\dot{\textbf{r}}^2_n -\phi + \mathbf{A}\cdot \dot{\mathbf{r}}_n \label{singlePartLagran}
\end{equation}

From electromagnetic theory, we borrow the ideas of set scalar field $\phi$ depend only on fixed source contributions $e.i$
\begin{equation}
\phi \equiv \phi(\mathbf{r}_1, \mathbf{r}_2, ..)
\end{equation}

We also presume the Gauge invariance of these fields $e.i$, and they can be chosen to satisfy the Lorentz condition:
\begin{equation}
\nabla \mathbf{A} + \frac{1}{c^2} \frac{\partial \phi}{\partial t} =0
\end{equation}

The preference for the form of potential energy like \ref{PotentialEnergyForm} is given mostly because:
\begin{itemize}
\item It provides a verified form for a self-consistent field that can exist even without the presence of sources, that in our case might correspond to the case of the value of the mass of all particle being zero.
\item It has a well tested covariant form for $U=A^\nu \dot{x}_\nu$ and $\partial_\nu A^\nu=0$ \cite{jackson.ElectroDynamics}.
\end{itemize}
\end{description}
\subsection{The equation of motion for the single particle in Cartesian coordinates \label{SinglePartEqMotionCart}} 
We are in the position now, after established previous assumptions for mass and field functions, to find the expression of the forces acting on every particle as one particle system, in Cartesian coordinates. For example, the inertial and net applied force in Cartesian coordinate on the $x$ direction for $n$ particle under the action of the external field is found from the Lagrange equation
\begin{equation}
 \dot{P}_{n,x}^{(1)} - Q_{n,x}^{(1)} = \mathcal{L}_{n,x} L_{{sp}_n} \label{genP1},
\end{equation}
where the external influence of the particles over the particle $n$ is on fields  $\mathbf{A}$ and $\phi$. Using the Lagrangian \ref{singlePartLagran} the inertial and net applied force in the $x$-direction have the form:
\begin{align}
& \dot{P}_{n,x}^{(1)} - Q_{n,x}^{(1)} = \mathcal{L}_{n,x} L_{{sp}_n} 
\nonumber \\ 
&\equiv \Big[ \frac{d}{dt}\frac{\partial}{\partial \dot{x}_n} 
- \frac{\partial}{\partial x_n}\Big](\frac{1}{2}m_n\dot{\textbf{r}}^2_n 
-\phi + \mathbf{A}\cdot \dot{\mathbf{r}}_n ) 
\nonumber \\
& =\frac{d}{dt}\Big(m_n \dot{x}_n 
+ \dot{\mathbf{r}}_n\frac{\partial \mathbf{A}}{\partial \dot{x}_n} + A_x \Big) 
- \frac{1}{2}\frac{\partial m_n}{\partial x_n}\dot{\textbf{r}}^2_n 
\nonumber \\
& \quad + \frac{\partial \phi}{\partial x_n}  
- \dot{\mathbf{r}}_n\frac{\partial \mathbf{A}}{\partial x_n}  
\nonumber \\ 
& =m_n\ddot{x}_n + (\vec{\nabla}_n m_n\cdot \dot{\mathbf{r}}_n) \dot{x}_n 
+ \ddot{\mathbf{r}}_n\frac{\partial \mathbf{A}}{\partial \dot{x}} 
\nonumber \\
& \quad+ \dot{\mathbf{r}}_n \frac{d}{dt}\Big(\frac{\partial \mathbf{A}}{\partial \dot{x}_n}\Big)
+ (\vec{\nabla}_{r_n}  A_x) \dot{\textbf{r}}_{n} 
+ (\vec{\nabla}_{\dot{r}_n}  A_x) \ddot{\textbf{r}}_{n} 
\nonumber \\
& \quad- \frac{1}{2}\frac{\partial m_n}{\partial x}\dot{\textbf{r}}^2_n 
+  \frac{\partial \phi}{\partial x_n}  
-  \dot{\mathbf{r}}_n\frac{\partial \mathbf{A}}{\partial x_n} \label{genP} 
\end{align}

The obtained inertial and net applied forces show its dependency with the derivatives of functions of masses and field with the coordinates and velocities. These derivatives functions are the new set of unknown variables of the system which must be found as the solution of the set of equations.
 
\section{New variables. The theory of Special Relativity on \texorpdfstring{$n$}-VMVF systems.}

The number of variables the system has increased. We must include then a new set of equations, so the equation system remains solvable. So far the set of equation of motion in eq. \ref{genP1} was settled using the Cartesian coordinates $r(x,y,z)$. This set of coordinates constitute a plane space, and they are related to the conservation of linear momentum which reflect the homogeneity in space. Others coordinates system like cylindrical, spherical, and elliptical won't increase the number of the independent equation because they are related to Cartesian coordinates through the transformation relations. 

The set of independent Lagrange equations must come from a set of coordinates also independent from Cartesian's which lead to an independent conservation law mirroring a different property of the space.

We propose the set of rotation angles  $r(\theta,\phi,\chi)$, used in the description of the rotation of physical systems, as the new independent set of coordinates needed for increasing the number of equations of the system. The Lagrange equations depending on the rotation angles lead to the law of the conservation of angular momentum in the absence of external torques which is related to the isotropy of space.

On the other side, the situation does not get better if the angles are also added as three more variables to the problem. This issue can be overcome by representing the position of the particle in both sets of coordinates. However, the set of angular coordinates forms a curved space and is independent of the Cartesian set of coordinates. It is not possible transforming the position vector expressed in Cartesian coordinates $r(x,y,z)$ into a vector expressed by the set of angles $r(\theta,\phi,\chi)$. There is a mathematical obstruction in that transformation. Nevertheless, the relation can be set by increasing one dimension to the 3-dimensional Euclidean space and setting the angular coordinates plus the radius as the spherical coordinate system of the $4$-dimensional Euclidean space.

We increase, then, the position vector in Cartesian coordinates $r(x,y,z)$ into a 4-vector, represented in its contravariant form by $r(x^0,x^1,x^2,x^3)$, whose inner product with its covariant form (\textit{1-form}) remains unchanged in all inertial frames of reference, imposing the condition:
\begin{equation}
x^2+y^2+z^2+w^2=R^2. \label{relatRelation}
\end{equation}
$R$ is the radius and is set as constant entering to the set of equations as a new parameter. The constant value for $R$ makes both sets of coordinates $\{x,y,z\}$ and $\{\theta,\phi,\chi\}$ independent. The transformation equations between the two set of variables are:
\begin{align}
x\equiv x^1&=R \sin \theta \sin \phi \sin \chi, \nonumber \\
y\equiv x^2&=R \sin \theta \sin \phi \cos \chi, \nonumber \\
z\equiv x^3&=R \sin \theta \cos \phi, \nonumber \\
w\equiv x^0&=R \cos \theta. \label{PosAngleRelation}
\end{align}
The parameter $R$ should be considered not related to each particle itself but the entire system. The computation of its value should be computed on future works.

The equation \ref{relatRelation} is the well known Lorentz condition. Unexpectedly, we start developing the theory under the Newtonian approach planning to extend the final result to its invariant form as every ``good'' theory should be. However, the need for describing the system using the 3-D angular coordinates to increase the number of independent equations, force the inclusion of relativistic approach. Under this point of view, the Lorentz condition in the space-time defines the 4-D surface which sets the 3-D space for the angular variables for all particles. We can say that the 3-D space of the angular coordinates is the stereographic projection of the 4-D sphere given by the Lorentz condition in the space-time.

Based on all the knowledge acquired from the physics community; we assume the modern approach of treating all coordinates as part of a non-Euclidean space and introduce the Theory of Relativity in our problem. It is well known the relation of the 4-component with time like $x^0 = \imagi c t$. Nevertheless the experimental background, the value $c$ must be constant since it can not depend on the others variables and must also be the same from any inertial frame. However, in the absence of mass and field's variation, the energy of the system must equal the Einstein formula $E = \sum_i m_i x^\nu x_\nu =\sum_i m_i c^2$. So, it is consistent that $c$ is the speed of light. So, from now on, we develop the theory under the Lorentz approach and apply the Minkowski algebra. 

The theory must now be extended to its relativistic formulation. We consider that:
\begin{itemize}
\item\item
We assume the postulates of the Theory of Relativity which state that 
\begin{quote}
``The laws of physics are the same for all observers in uniform motion relative to one another''. 
\end{quote}
and
\begin{quote}
``The speed of light in a vacuum is the same for all observers, regardless of their relative motion or of the motion of the light source.''
\end{quote}
\item 
In the present theory, the four components of the vector position $x^\nu$ are not independent due to  Lorentz constraint. However, we follow Dirac point of view, from where such constraint is treated as how he described: ``a weak condition'' \cite{goldstein321}, what means that the constraint should be imposed after all derivation processes have been carrying through.
\item 
The scalar and vector fields, $\phi$ and $\mathbf{A}$ are now replaced by the four-dimensional field $A^\nu$. The proposed relativistic Lagrangian, $L_n$, for a single particle under the action of an external field, now has the form:
\begin{equation}
L_n=\frac{1}{2}m_n \dot{r}^\nu_n \dot{r}_{n,\nu} -A^\nu \dot{r}_{n,\nu} \label{RelLagrangian}
\end{equation}

\item One of the invariants against a Lorentz transformation is the infinitesimal square of the distance in the Minkowski space,
\begin{equation}
ds^2 = ds^{'2} = dx_0^2+dx_1^2+dx_2^2+dx_3^2.
\end{equation}
For establishing the relations for the differentiation of any vector in the spacetime, it was defined as the invariant quantity known as the proper time of the system. By the inclusion of this concept to our problem, the covariant Lagrange equations evolve now, not on time, but on the proper time or $\tau$. 

\item The covariant Lagrange operator depending on the Lorentzian coordinates now is:
\begin{equation}
\mathcal{L}_{n,\nu}  \equiv \Big[ \frac{d}{d\tau}\frac{\partial}{\partial \dot{x}^\nu_n} - \frac{\partial}{\partial x_n^\nu}\Big], \;\; x^\nu=\{x^0,x^1,x^2,x^3\} \label{lagrangetranslation}
\end{equation}
while the same operator depending on the angular coordinates is
\begin{equation}
\mathcal{L}_{n,\xi}  \equiv \Big[ \frac{d}{d\tau}\frac{\partial}{\partial \dot{\xi}_n} - \frac{\partial}{\partial \xi_n}\Big], \;\; \xi=\{ \theta,\phi,\chi\}. \label{lagrangeRotation}
\end{equation}
\item The relativistic formulation of the present approach, along with the extra dimension, includes new variables to the description of the problem related to this new coordinate. However, they must not increase internal degrees of freedom of the system. These quantities must be constrained in somehow with the already existing variables in the non-relativistic picture. The new variables are 
\begin{equation}
\frac{\partial m_n}{\partial x_{n}^0}, \;\;\; \frac{\partial m_n}{\partial \dot{x}_{n}^0}, \;\;\; \frac{\partial A^0}{\partial x^\nu_{n}} , \;\;\; \frac{\partial A^0}{\partial \dot{x}^\nu_{n}}, \;\;\; \frac{\partial A^i}{\partial x^0_n}, \;\;\; \frac{\partial A^i}{\partial \dot{x}^0_{n}}.\label{relConstraint}
\end{equation}
\item
Considering the four coordinate of the field $A^0$ as the scalar field, the approaches from section  \ref{classicalAssump1} in the relativistic frame, have the form:
\begin{align}
\frac{\partial A^0}{\partial \dot{x}^\nu_{n}}=0, \quad
\partial_\nu A^\nu = 0 \;\forall\; n, \label{relConstraint1}
\end{align}
remaining only the constraints for the variables related to the variation of the mass.

\item We can classify the variations of the mass into two types: the ``structural'' variations and the inertial's. The first type includes the ones introduced in the non-relativistic approach. Their non zero derivatives are $\frac{\partial m_n}{\partial x^i_n}$ with $\{x^1_n,x^2_n,x^3_n\} \equiv \{x,y,z\}$, while the inertial variation, which is related to the inertial frames of references, is added because of the relativistic theory. It includes the derivative $\frac{\partial m_n}{\partial x^0_n}$ as a new degree of freedom. 

According the relativistic theory, the variation of the mass of the particle depends on the velocity of the particle as obtained from the relativistic momentum:
\begin{equation*}
m = \frac{m_0}{\sqrt{1 - \frac{v^2}{c^2}}}.
\end{equation*}
Then we can consider the derivative $\frac{\partial m_n}{\partial x^0_n}=0$  and $\frac{\partial m_n}{\partial \dot{x}^\nu_n} \neq 0$.
If the invariant Lagrangian of equation \ref{RelLagrangian} is applied in case of the isolated free particle, the equation of motion has the form:
\begin{align}
\mathcal{L}_{\mu} L &\equiv 
\Big[ \frac{d}{d\tau}\frac{\partial}{\partial \dot{x}^\mu} 
- \frac{\partial}{\partial x^\mu}\Big] \Big( \frac{1}{2}m\dot{x}^\nu\dot{x}_\nu \Big)=0 
\nonumber \\
& = \frac{d}{d\tau} (m\dot{x}_\mu) 
= \dot{m}\dot{x}_\mu + m\ddot{x}_\mu =0. \nonumber
\end{align}
Replacing $\dot{m} = \frac{\partial m}{\partial \dot{x}^\mu} \ddot{x}^\mu$, we obtain $1n$ + $4n$ = $5n$  constraints:
\begin{align}
\frac{\partial m_n}{\partial x^0_n}=0, \quad 
\frac{\partial m_n}{\partial \dot{x}^\nu_n} \ddot{x}^\nu_n \dot{x}_{n;\mu} + m_n \ddot{x}_{n;\mu} =0 \label{mass0compConstraint}
\end{align}
which entirely determine the inertial variation and left unchanged the number of degrees of freedom of the system.
\end{itemize}
The choice of setting the field and the mass of the particle as unknown quantities while the metric tensor remains constant to establish a different point of view from theories like General Relativity, where the form of the field is fixed, and it is the space which is modified in the presence of a field or a massive object. Also, as the field is proposed as the unique physical object that connects particles, it includes all possible fundamental interactions and coupling. Under this approach, as the form of the field is found as part of the solution of the problem, the presence or the variation of the mass of a massive object is traduced directly in a variation of all the fundamental interactions.

\subsection{The equation of motion for the single particle}
We update now the inertial and net applied forces acting on a particle assuming the one particle system under the action of an external field as exposed on section \ref{SinglePartEqMotionCart} using the new sets of coordinates: the Lorentzian and the angular's, the new assumptions and replacing the time by the proper time.

\subsubsection{The equation of motion for the single particle in Lorentzian}
The $\mu$ component of the inertial and applied forces acting on particle $n$ is
\begin{align*}
&\dot{P}_{n,\mu}^{(1)} - Q_{n,\mu}^{(1)} 
= \mathcal{L}_{n,\mu} L_{{sp}_n}
\\
&= \mathcal{L}_{n,\mu}(\frac{1}{2}m_n \dot{x}^\nu_n \dot{x}_{n;\nu} -A^\nu \dot{x}_{n;\nu} )  \\
&= \Big[ \frac{d}{d\tau}\frac{\partial}{\partial \dot{x}_n^\mu} 
- \frac{\partial}{\partial x_n^\mu}\Big](\frac{1}{2}m_n \dot{x}^\nu_n \dot{x}_{n;\nu} 
-A^\nu \dot{x}_{n;\nu} ) \\
& =\frac{d}{d\tau}\Big[m_n \dot{x}_{n;\mu} 
- \Big(\frac{\partial A^\nu}{\partial \dot{x}_n^\mu}\Big) \dot{x}_n^\nu 
- A_\mu \Big]
\\
&- \Big[ \frac{1}{2}\frac{\partial m_n}{\partial x_n^\mu}\dot{x}^\nu_n \dot{x}_{n;\nu} 
 -  \Big(\frac{\partial A^\nu}{\partial x_n^\mu}\Big)\dot{x}_{n;\nu}  \Big]
\end{align*}
or
\begin{align}
\dot{P}_{n,\mu}^{(1)} &- Q_{n,\mu}^{(1)} =m_n \ddot{x}_{n;\mu} 
+ \Big[\square_n^\nu m_n \dot{x}_{n;\nu}\Big] \dot{x}_{n;\mu} 
\nonumber\\ 
&+ \Big[\square_n^{\dot{\nu}} m_n \ddot{x}_{n;\nu}\Big] \dot{x}_{n;\mu} 
- \Big(\frac{\partial A^\nu}{\partial \dot{x}_n^\mu}\Big)\ddot{x}_{n;\nu} 
\nonumber\\ 
&- \frac{d}{d\tau}\Big(\frac{\partial A^\nu}{\partial \dot{x}_n^\mu} \Big) \dot{x}_{n;\nu}
- \frac{\partial A_\mu}{\partial x_n^\nu}\dot{x}_n^\nu 
- \frac{\partial A_\mu}{\partial \dot{x}_n^\nu}\ddot{x}_n^\nu  
\nonumber\\ 
&- \frac{1}{2}\frac{\partial m_n}{\partial x_n^\mu}\dot{x}^\nu_n \dot{x}_{n;\nu} 
+ \Big(\frac{\partial A^\nu}{\partial x_n^\mu}\Big)\dot{x}_{n;\nu} 
\label{genP1Rel}.
\end{align}
%\begin{equation}
%& =\Big[m_n(0)  + \square_n^\nu m_n x_{n;\nu} + \frac{1}{2}{\square_n^\nu}^2 m_n  (x_{n;\nu})^2...\Big]\ddot{x}_{n;\mu} + \Big[\square_n^\nu m_n \dot{x}_{n;\nu}\Big] \dot{x}_{n;\mu} \nonumber\\ 
%&\;\;\;\;\;\; - \Big(\frac{\partial A^\nu}{\partial \dot{x}_n^\mu}\Big)\ddot{x}_{n;\nu}
%- \frac{d}{d\tau}\Big(\frac{\partial A^\nu}{\partial \dot{x}_n^\mu} \Big) \dot{x}_{n;\nu}- \Big[ \frac{\partial A_\mu}{\partial x_n^\nu}\dot{x}_n^\nu + \frac{\partial A_\mu}{\partial \dot{x}_n^\nu}\ddot{x}_n^\nu \Big] \nonumber \\ &\;\;\;\;\;\; - \frac{1}{2}\frac{\partial m_n}{\partial x_n^\mu}\dot{x}^\nu_n \dot{x}_{n;\nu} + \Big(\frac{\partial A^\nu}{\partial x_n^\mu}\Big)\dot{x}_{n;\nu. 
%\end{equation}
\subsubsection{The equation of motion for the single particle in angular coordinates}
It is useful to obtain some relations before we compute the net applied, and inertial forces applied for the angular's coordinates. 

From Goldstein \citep{goldstein}, we get the equations:
\begin{equation}
 \frac{d}{dt}\Big(\frac{\partial x_l}{\partial q_i}\Big)= \frac{\partial \dot{x}_l}{\partial q_i} \;\; \text{ and } \;\; \frac{\partial \dot{x}_l}{\partial \dot{q}_i} = \frac{\partial x_l}{\partial q_i}, \label{relations0}
\end{equation}
Using the compound derivative law, we obtain can other relations up to the second order derivatives, which we compute for later uses:
\begin{align}
\dot{x_l} &= \sum_i \frac{\partial x_l}{\partial q_i} \dot{q}_i + \frac{\partial x_l}{\partial t}
\nonumber\\
\ddot{x_l} &= \frac{d}{dt}\Big( \sum_i \frac{\partial x_l}{\partial q_i} \dot{q}_i + \frac{\partial x_l}{\partial t} \Big)
\nonumber\\
&= \sum_{i}\Big[ \sum_j \frac{\partial^2 x_l}{\partial q_i \partial q_j} \dot{q}_i \dot{q}_j + \frac{\partial x_l}{\partial q_i} \ddot{q}_i + 2\frac{\partial^2 x_l}{\partial q_i \partial t} \dot{q}_i \Big]
\nonumber\\
& \qquad+ \frac{\partial^2 x_l}{\partial t^2}. \label{relations1}
\end{align}
On the other side, deriving $\dot{x_l}$ and  $\ddot{x_l}$ we obtain
\begin{align}
\frac{\partial \dot{x_l}}{\partial q_k} &= \sum_i \frac{\partial^2 x_l}{\partial q_i \partial q_k} \dot{q}_i + \frac{\partial^2 x_l}{\partial t \partial q_k}
\nonumber\\
\frac{\partial \ddot{x_l}}{\partial q_k}&= \sum_{i}\Big[ \sum_j \frac{\partial^3 x_l}{\partial q_i \partial q_j \partial q_k} \dot{q}_i \dot{q}_j + \frac{\partial^2 x_l}{\partial q_i \partial q_k} \ddot{q}_i 
\nonumber\\
&\quad+ 2\frac{\partial^3 x_l}{\partial q_i \partial q_k \partial t} \dot{q}_i \Big] + \frac{\partial^3 x_l}{ \partial t^2 \partial q_k}  
\nonumber\\
\frac{\partial \ddot{x_l}}{\partial \dot{q}_k} &= \sum_{i}\Big[ \sum_j \frac{\partial^2 x_l}{\partial q_i \partial q_j} (\dot{q}_i \delta_{jk} + \dot{q}_j \delta_{ik}) 
\nonumber\\ 
&+  2\frac{\partial^2 x_l}{\partial q_i \partial t} \delta_{ik}
\Big] 
\nonumber\\
&=   \sum_i 2\frac{\partial^2 x_l}{\partial q_i \partial q_k} \dot{q}_i + 2\frac{\partial^2 x_l}{\partial t \partial q_k} = 2 \frac{\partial \dot{x_l}}{\partial q_k}
\nonumber\\
\frac{\partial \ddot{x_l}}{\partial \ddot{q}_k} &= \sum_i \frac{\partial x_l}{\partial q_i} \delta_{ik} \;\;=\frac{\partial x_l}{\partial q_k}. \label{relations2}
\end{align}
Also
\begin{align}
&\frac{d}{dt}\Big( \frac{\partial \dot{x_l}}{\partial q_k} \Big) = \frac{d}{dt}\Big(  \sum_i \frac{\partial^2 x_l}{\partial q_i \partial q_k} \dot{q}_i + \frac{\partial^2 x_l}{\partial t \partial q_k} \Big)
\nonumber\\
&= \sum_{i}\Big[ \sum_j \frac{\partial^3 x_l}{\partial q_i \partial q_j \partial q_k} \dot{q}_i \dot{q}_j 
+ \frac{\partial^2 x_l}{\partial q_i \partial q_k} \ddot{q}_i 
\nonumber\\
&+ 2\frac{\partial^3 x_l}{\partial q_i \partial q_k \partial t} \dot{q}_i \Big] 
+ \frac{\partial^3 x_l}{ \partial t^2 \partial q_k}. \label{relations3}
\end{align}
Comparing \ref{relations2} and \ref{relations3}, we get the relation
\begin{equation}
\frac{d}{dt}\Big( \frac{\partial \dot{x_l}}{\partial q_k} \Big)= \frac{\partial \ddot{x_l}}{\partial q_k} \;\;\;\; \text{or} \;\;\;\; \frac{d^2}{dt^2}\Big( \frac{\partial x_l}{\partial q_k} \Big) = \frac{\partial \ddot{x_l}}{\partial q_k} \label{relations4}.
\end{equation}
Applying the previous results to our coordinates systems and using the compound derivative law up to the second order, the derivative on angular variables can be replaced as:
\begin{align}
\frac{\partial}{\partial \xi_i}&= \frac{\partial x^\nu}{\partial \xi_i} \frac{\partial}{\partial x^\nu} + \frac{\partial \dot{x}^\nu}{\partial \xi_i} \frac{\partial}{\partial \dot{x}^\nu} + \frac{\partial \ddot{x}^\nu}{\partial \xi_i} \frac{\partial}{\partial \ddot{x}^\nu}
\nonumber\\
&=\frac{\partial x^\nu}{\partial \xi_i} \frac{\partial}{\partial x^\nu} + \frac{d}{d\tau}\Big(\frac{\partial x^\nu}{\partial \xi_i}\Big) \frac{\partial}{\partial \dot{x}^\nu} 
\nonumber\\ 
&\quad+ \frac{d^2}{d\tau^2}\Big(\frac{\partial x^\nu}{\partial \xi_i}\Big) \frac{\partial}{\partial \ddot{x}^\nu} 
\nonumber\\
\frac{\partial}{\partial \dot{\xi}_i}&= \frac{\partial \dot{x}^\nu}{\partial \dot{\xi}_i} \frac{\partial}{\partial \dot{x}^\nu} + \frac{\partial \ddot{x}^\nu}{\partial \dot{\xi}_i} \frac{\partial}{\partial \ddot{x}^\nu} 
\nonumber\\
&= \frac{\partial x^\nu}{\partial \xi_i} \frac{\partial}{\partial \dot{x}^\nu} + 2\frac{d}{d\tau}\Big(\frac{\partial x^\nu}{\partial \xi_i}\Big) \frac{\partial}{\partial \ddot{x}^\nu}
\nonumber\\
\frac{\partial}{\partial \ddot{\xi}_i}&= \frac{\partial \ddot{x}^\nu}{\partial \ddot{\xi}_i} \frac{\partial}{\partial \ddot{x}^\nu}  \;\; = \frac{\partial x^\nu}{\partial \xi_i} \frac{\partial}{\partial \ddot{x}^\nu} \label{relations5}
\end{align}
In matrix notation, we have:
\begin{align}
\vec{\nabla}_\xi&= D_{\;\xi}^\nu \square_\nu + \frac{d }{d\tau}\Big(D_{\;\xi}^\nu \Big) \square_{\dot{\nu}} + \frac{d^2}{d\tau^2}\Big(D_{\;\xi}^\nu \Big) \square_{\ddot{\nu}} 
\nonumber\\
\vec{\nabla}_{\dot{\xi}}&= D_{\;\xi}^\nu \square_{\dot{\nu}} + 2\frac{d }{d\tau}\Big(D_{\;\xi}^\nu \Big) \square_{\ddot{\nu}}
\nonumber\\
\vec{\nabla}_{\ddot{\xi}}&= D_{\;\xi}^\nu \square_{\ddot{\nu}} \label{relationsMatrix}
\end{align}
where $D_{\;\xi}^\nu$ corresponds to the elements of the $3\times 4$ matrix of the derivatives $\frac{\partial x^\nu}{\partial \xi_i}$. The elements $D_{\;\xi}^\nu$ are obtained from the transformation between the sets $\{x^0,x^1,x^2,x^3\}$ and $\{\theta,\phi,\chi\}$ as showed in eq. \ref{PosAngleRelation}. From the definition, we have:
\begin{align}
\cos \theta &= x^0/R \\
\cos \phi &= x^3/(R\sin \theta)=	x^3/\sqrt{R^2-(x^0)^2}\\
\cos \chi &= x^2/(R\sin \theta \sin \phi)\nonumber\\
&=	x^2/\sqrt{R^2-(x^0)^2 - (x^3)^2}\\
\tan \chi &=x^1/x^2.
\end{align}
After some straightforward derivation, we obtain:
%\begin{widetext}
\begin{equation}
D_{\;\xi}^\nu = 
\begin{bmatrix}
\frac{x^1x^0}{\sqrt{R^2-({x^0})^2}} & \frac{x^2 x^0}{\sqrt{R^2-({x^0})^2}} & \frac{x^1x^3}{\sqrt{R^2-({x^0})^2}} & -\frac{\sqrt{R^2-({x^0})^2}}{R}\\
\frac{x^1x^3}{\sqrt{R^2-({x^0})^2-({x^3})^2}} & \frac{x^2x^3}{\sqrt{R^2-({x^0})^2-({x^3})^2}} & - {\scriptstyle \sqrt{R^2-({x^0})^2-({x^3})^2}} & 0 \\
x^2 & -x^1 & 0 & 0
\end{bmatrix}.
\end{equation}
%being $(x^\nu)^2\equiv x^\nu x_\nu$ where Einstein summation notation is was not apply. 
%\end{widetext}

We replace the components $\theta,\phi,\chi$ for particle $n$ by $\xi_{n,i}...\; i=(1,2,3)$ respectively. The $\xi_i$ component of the angular net and applied forces acting on particle $n$, known as ``torque'', assuming the one particle system and represented as $\dot{L}_{\;\xi_{n,i}}^{(1)} - T_{\;\xi_{n,i}}^{(1)}$, are:
\begin{align}
&\dot{L}_{\;\xi_{n,i}}^{(1)} - T_{\;\xi_{n,i}}^{(1)} 
= \mathcal{L}_{\xi_{n,i}} L_{{sp}_n}
\nonumber \\
&\equiv \Big[ \frac{d}{d\tau}\frac{\partial}{\partial \dot{\xi}_{n,i}} - \frac{\partial}{\partial \xi_{n,i}}\Big]
(\frac{1}{2}m_n \dot{x}^\nu_n \dot{x}_{n;\nu} -A^\nu \dot{x}_{n;\nu} ).
\end{align}
Replacing the results of equations \ref{relationsMatrix}:
\begin{align}
&\dot{L}_{\;\xi_{n,i}}^{(1)} - T_{\;\xi_{n,i}}^{(1)} =
\Big[ \frac{d}{d\tau}\Big(  D_{\;\xi_{n,i}}^\mu \square_{n,\dot{\mu}} \Big) 
- D_{\;\xi_{n,i}}^\mu \square_{n,\mu} 
\nonumber \\
&- \frac{d }{d\tau}\Big(D_{\;\xi_{n,i}}^\mu \Big) \square_{n,\dot{\mu}}\Big]
 L_{{sp}_n} 
\nonumber \\
&= D_{\;\xi_{n,i}}^\mu \Big[  \frac{d}{d\tau}\Big(  \square_{n,\dot{\mu}} \Big) - \square_{n,\mu} \Big] L_{{sp}_n} \nonumber \\
&\dot{L}_{\;\xi_{n,i}}^{(1)} - T_{\;\xi_{n,i}}^{(1)} =  D_{\;\xi_{n,i}}^\mu \Big( \dot{P}_{n,\mu}^{(1)} - Q_{n,\mu}^{(1)} \Big).  \label{genLRel}
\end{align}
An example of the previous result, the angular component $\xi_{n,3}\equiv \chi$ of the net applied and inertial torque has the expression:
\begin{align}
&\dot{L}_{\;\xi_{n,3}}^{(1)} - T_{\;\xi_{n,3}}^{(1)} = \mathcal{L}_{\xi_{n,3}}(\frac{1}{2}m_n \dot{x}^\nu_n \dot{x}_{n;\nu} -A^\nu \dot{x}_{n;\nu} )  
\nonumber \\
&= D_{\;\xi_{n,3}}^\mu \Big( \dot{P}_{n,\mu}^{(1)} - Q_{n,\mu}^{(1)} \Big) 
\nonumber \\
&= 
x^2 \Big( \dot{P}_{n,1}^{(1)} - Q_{n,1}^{(1)} \Big)
-x^1 \Big( \dot{P}_{n,2}^{(1)} - Q_{n,2}^{(1)} \Big),
\end{align}
which is the well-known expression in tree dimensions $\mathbf{T}=\mathbf{r} \times \mathbf{F}$.

\section{Obtaining the equations of motion for \texorpdfstring{$n$}-VMVF systems}

The Lagrangian of $n$-VMVF systems is unknown at this point. However, from the extension of the D'Alembert principle, we know that the solution of the Lagrange equation for the $n$-VMVF systems is the sum of the inertial and the net applied forces on all particles. On the other side, under the assumption from section \ref{genForceApproach} of the approaching of the isolated $n$- particle system to $n$ systems of one particle Eq. \ref{ApproachAppliedForces}, we can obtain those forces as the net forces that act over each particle under the action of the external field formed by the rest of the particle for both coordinates as shown in equations \ref{genP1Rel} and \ref{genLRel}. 

%Starting from an initial Lagrange function, if we compare the obtained forces on each coordinate system, we can obtain then a set of relations that will constrain the motion of the particles of the system. 
In summary, our set of equations using the Lorentzian coordinates is 
\begin{align}
&\mathcal{L}_{n,\mu}L_{sys} =0  \label{partSysEq1} \\ 
&\mathcal{L}_{n,\mu}L_{sys}= \sum_{n'} \dot{P}_{{n'},\mu}^{(N)} - Q_{{n'},\mu}^{(N)}\label{partSysEq2}\\
&\sum_n \dot{P}_{{n'},\mu}^{(N)} - Q_{{n'},\mu}^{(N)}  = \sum_n \dot{P}_{{n'},\mu}^{(1)} - Q_{{n'},\mu}^{(1)}, \label{ApproachAppliedForcesLorentz}
\end{align}
and the set of equations using the angular coordinates is
\begin{align}
&\mathcal{L}_{\xi_{n,i}}L_{sys} =0 \label{partSysEq3} \\
&\mathcal{L}_{\xi_{n,i}}L_{sys}= \sum_{n'}\dot{L}_{\;\xi_{{n'},i}}^{(N)} - T_{\;\xi_{{n'},i}}^{(N)}
\label{partSysEq4}\\
&\sum_n \dot{L}_{\;\xi_{{n'},i}}^{(N)} - T_{\;\xi_{{n'},i}}^{(N)} = \sum_n \dot{L}_{\;\xi_{{n'},i}}^{(1)} - T_{\;\xi_{{n'},i}}^{(1)}. \label{ApproachAppliedForcesAngular} 
\end{align}

The derivatives of masses and the field are unknown quantities of the equation system; however, they are not generalized coordinates of the variational problem. This type of solution, which involves that kind of treatment for a group of variables and includes two sets of constrained set of Lagrange equations, is new as a solution of a classical problem, at least to the best of our knowledge.

Recalling the equation \ref{splittedLagEq}, the total Lagrangian for particle $n$-VMVF systems can be divided into two terms
\begin{equation*}
L_{sys} = L_{[\dot{m}_n= 0]} + L_{[\dot{m}_n\neq 0]} = \sum_{n'=1}^N L_{sp_{n'}} + L_{[\dot{m}_n\neq 0]},
\end{equation*}
where the first term considers constant masses and is equal to the well-known Lagrangian for a isolated particle system $L_{sp}$:
\begin{equation*}
L_{sp} = \sum_{n'} L_{sp_{n'}} = \sum_{n'} \frac{1}{2}m_{n'} \dot{x}^\nu_{n'} \dot{x}_{n';\nu}.
\end{equation*}
When the mass variation is taking into account, the field is included and the Lagrange function take the form
\begin{equation}
L_{sys} = \sum_{n'} \frac{1}{2}m_{n'} \dot{x}^\nu_{n'} \dot{x}_{n';\nu} -A^\nu \dot{x}_{n';\nu}. \label{nonIntPartSys}
\end{equation}

The strategy to follow for obtaining the final Lagrangian for $n$-VMVF systems is to propose the initial Lagrange function $L_{sys}$ and obtain the net applied and generalized inertial forces. The $n$-degeneracy of the Lagrange equations lead to $n$ independent equations by the comparison of the obtained equations with the net applied and generalized inertial forces under the one particle system approach shown on previous sections. Those constraints should be included in the final Lagrangian using the Lagrange multiplier method.

\subsection{Constraints for \texorpdfstring{$n$}-VMVF systems in Lorentzian coordinates}
The equation of motion of particle $n$ in the $\mu$ direction as a constituent of an isolated $n$-VMVF system using the proposed Lagrangian is
\begin{align*}
&\mathcal{L}_{n,\mu}( L_{sys}) 
= \mathcal{L}_{n,\mu}( \sum_{n'} \frac{1}{2}m_{n'} \dot{x}^\nu_{n'} \dot{x}_{n';\nu} -A^\nu \dot{x}_{n';\nu} )
\\
&= \Big[ \frac{d}{d\tau}\frac{\partial}{\partial \dot{x}_n^\mu} - \frac{\partial}{\partial x_n^\mu}\Big]( \sum_{n'} \frac{1}{2}m_{n'} \dot{x}^\nu_{n'} \dot{x}_{{n'};\nu} -A^\nu \dot{x}_{{n'};\nu}. )
\\
&=0
\end{align*}
We consider now all particles as parts of the system, and because of that, the coordinates and velocities of all particles are now treated as variables of the system. We divide, for simplicity, the Lagrange's function into one term related to the particle $n$ and another grouping the terms of the rest of the particles
\begin{equation}
\mathcal{L}_{n,\mu}( \sum_{n'} L_{n'})= \mathcal{L}_{n,\mu}( L_n + \sum_{{n'} \neq n} L_{n'}).\label{lagranDivision1}
\end{equation}
The first term, using equation \ref{genP1Rel}, have the form:
\begin{align}
 \mathcal{L}_{n,\mu} L_n&=\Big[ \frac{d}{d\tau}\frac{\partial}{\partial \dot{x}_n^\mu} 
- \frac{\partial}{\partial x_n^\mu}\Big](\frac{1}{2}m_n \dot{x}^\nu_n \dot{x}_{n;\nu} -A^\nu \dot{x}_{n;\nu} )
\nonumber \\
&=\frac{d}{d\tau}\Big[m_n \dot{x}_{n;\mu} 
- \Big(\frac{\partial A^\nu}{\partial \dot{x}_n^\mu}\Big) \dot{x}_{n;\mu} 
- A_\mu \Big] 
\nonumber \\
&\quad - \Big[ \frac{1}{2}\frac{\partial m_n}{\partial x_n^\mu}\dot{x}^\nu_n \dot{x}_{n;\nu} 
-  \Big(\frac{\partial A^\nu}{\partial x_n^\mu}\Big)\dot{x}_{n;\mu}  \Big] 
\nonumber \\
&= \dot{P}_{n,\mu}^{(1)} - Q_{n,\mu}^{(1)} - \sum_{l \neq n} \Big( \frac{\partial A_\mu}{\partial x_{l}^\nu}\dot{x}_l^\nu + \frac{\partial A_\mu}{\partial \dot{x}_l^\nu}\ddot{x}_l^\nu \Big), \label{TransConstraintEq0} 
\end{align}
where we have replaced the equation \ref{genP1Rel} in the result. 

The only non-null terms in the last part of motion equation \ref{lagranDivision1} is related to potential energy since terms $\frac{1}{2}m_n \dot{x}^\nu_n \dot{x}_{n;\nu}$ depends only on particle $n$ coordinates. In that case, we obtain:
\begin{align}
&\mathcal{L}_{n,\mu}(\sum_{{n'} \neq n} L_{n'}) = \mathcal{L}_{n,\mu}(\sum_{{n'} \neq n} -A^\nu \dot{x}_{{n'};\nu})
\nonumber \\
&= - \sum_{{n'} \neq n} \Big\{ \frac{d}{d\tau}\Big[ \Big(\frac{\partial A^\nu}{\partial \dot{x}_{n}^\mu}\Big) \dot{x}_{n';\nu} \Big] - \Big(\frac{\partial A^\nu}{\partial x_{n}^\mu}\Big)\dot{x}_{n';\nu} \Big\}. \label{TransConstraintEq1}
\end{align}
Putting all together, the motion equation of particle $n$ in the $\mu$ direction is
\begin{align}
\mathcal{L}_{n,\mu}L_{sys} &= \dot{P}_{n,\mu}^{(1)} - Q_{n,\mu}^{(1)} 
\nonumber \\ 
&- \sum_{{n'} \neq n} \Big\{ \frac{d}{d\tau}\Big[ \Big(\frac{\partial A^\nu}{\partial \dot{x}_{n}^\mu}\Big) \dot{x}_{n';\nu} \Big] 
- \Big(\frac{\partial A^\nu}{\partial x_{n}^\mu}\Big)\dot{x}_{n';\nu}  
\nonumber \\ 
& + \frac{\partial A_\mu}{\partial x_{n'}^\nu}\dot{x}_{n'}^\nu + \frac{\partial A_\mu}{\partial \dot{x}_{n'}^\nu}\ddot{x}_{n'}^\nu \Big\}=0.
\end{align}

The comparison of the equations given by the extension of D'Alembert principle and the approximation of $n$ systems of one particle:
\begin{equation}
\mathcal{L}_{n,\mu}( L_{sys})
= \sum_{n'} \dot{P}_{{n'},\mu}^{(N)} - Q_{{n'},\mu}^{(N)} 
= \sum_{n'} \dot{P}_{{n'},\mu}^{(1)} - Q_{{n'},\mu}^{(1)},
\end{equation}
results in the following 3-$n$ independent equation:
\begin{align}
\Phi_{\mu_n}& = \sum_{{n'} \neq n}  \Big[
\Big(\square_{n'}^\alpha m_{n'} \dot{x}_{n';\alpha}\Big)g^\nu_\mu
- \frac{1}{2}\frac{\partial m_{n'}}{\partial x_{n'}^\mu}\dot{x}^\nu_{n'}
\nonumber \\
&+ \frac{d}{d\tau} \Big(\frac{\partial A^\nu}{\partial \dot{x}_{n}^\mu}
- \frac{\partial A^\nu}{\partial \dot{x}_{n'}^\mu}
\Big) 
+ \frac{\partial A^\nu}{\partial x_{n'}^\mu}
- \frac{\partial A^\nu}{\partial x_{n}^\mu}
\Big]\dot{x}_{n';\nu}
\nonumber \\
&+ \Big[ 
m_{n'}g^\nu_\mu + \frac{\partial A^\nu}{\partial x_{n}^\mu}
- \frac{\partial A^\nu}{\partial x_{n'}^\mu}
+ \frac{\partial A_\mu}{\partial x_{n'}^\nu}
\Big] \ddot{x}_{n';\nu}
\nonumber \\
&+ \square_{n'}^{\dot{\mu}} m_{n'} \ddot{x}_{n';\mu}  \dot{x}_{n';\nu}.
\label{TransConstraintEq}
\end{align}
Together to the least action principle equation 
\begin{equation}
\mathcal{L}_{n,\mu}( L_{sys}) = \mathcal{L}_{n,\mu}( \sum_{n'} \frac{1}{2}m_{n'} \dot{x}^\nu_{n'} \dot{x}_{n';\nu} -A^\nu \dot{x}_{n';\nu} ) =0 \label{TransLeastActionEq} 
\end{equation}
we have the 6-$n$ independent equations needed to solve the problem. Actually, there are 8-$n$ equations, however, the Lorentz's constraints must reduce them to the final 6-$n$ independent equations. 

%The constraint equations depend on variables
%\begin{equation}
%\{x_n^\nu\},\{\dot{x}_n^\nu\},\{\ddot{x}_n^\nu\}
%,\{\frac{\partial m_n}{\partial x_n^\mu }\}
%,\{\frac{\partial m_n}{\partial \dot{x}_n^\mu }\}, 
%\{\frac{\partial A^\nu}{\partial x_n^\mu }\}
%\{\frac{\partial A^\nu}{\partial \dot{x}_n^\mu }\}
%\label{ConstraintDep}
%\end{equation}
%which means we need more equations for the system being solvable. Thus, the inclusion of angular variables.

\subsection{Constraints for \texorpdfstring{$n$}-VMUF systems in angular coordinates}
The motion equation of particle $n$ rotating around axis in the $i$ direction for isolated $n$-VMVF systems using the starting Lagrangian is
\begin{align}
&\mathcal{L}_{\xi_{n,i}}( L_{sys})  =0 \nonumber \\
&=  D_{\;\xi_{n,i}}^\mu \Big[  \frac{d}{d\tau}\Big(  \square_{n,\dot{\mu}} \Big) - \square_{n,\mu} \Big]
(L_{sys}) 
\nonumber \\
&\equiv  D_{\;\xi_{n,i}}^\mu \Big[  \frac{d}{d\tau}\Big(  \square_{n,\dot{\mu}} \Big) 
- \square_{n,\mu} \Big] (L_n + \sum_{{n'} \neq n} L_{n'}),
\end{align}
where we divided the Lagrangian again like \ref{lagranDivision1}. Using the previous result on eq. \ref{TransConstraintEq0}, the first term in the sum is:
\begin{align}
\mathcal{L}_{\xi_{n,i}}( L_n) &= D_{\;\xi_{n,i}}^\mu \Big[  \frac{d}{d\tau}\Big(  \square_{n,\dot{\mu}} \Big) - \square_{n,\mu} \Big]L_n 
\nonumber \\ 
& = D_{\;\xi_{n,i}}^\mu \Big[ \dot{P}_{n,\mu}^{(1)} - Q_{n,\mu}^{(1)} 
\nonumber \\ 
&- \sum_{n' \neq n} \Big( \frac{\partial A_\mu}{\partial x_{n'}^\nu}\dot{x}_{n'}^\nu 
+ \frac{\partial A_\mu}{\partial \dot{x}_{n'}^\nu}\ddot{x}_{n'}^\nu\Big) \Big].
\end{align}
The only nonnull terms of the last part are related to the field's addends as eq. \ref{TransConstraintEq1}. In that case, we obtain:
\begin{align}
\mathcal{L}_{\xi_{n,i}}(\sum_{{n'} \neq n} L_{n'})
%&=  D_{\;\xi_{n,i}}^\mu \Big[  \frac{d}{d\tau}\Big(  \square_{n,\dot{\mu}} \Big) - \square_{n,\mu} \Big] (\sum_{{n'} \neq n} L_{n'})
%\nonumber \\
&= - D_{\;\xi_{n,i}}^\mu \sum_{{n'} \neq n} \Big\{ \frac{d}{d\tau}\Big[ \Big(\frac{\partial A^\nu}{\partial \dot{x}_{n}^\mu}\Big) \dot{x}_{n';\nu} \Big]
\nonumber \\
&- \Big(\frac{\partial A^\nu}{\partial x_{n}^\mu}\Big)\dot{x}_{n';\nu} \Big\}.
\end{align}

The comparison of the equations given by the extension of D'Alembert principle and the approximation of $n$ systems of one particle under the action of the external field for the angular coordinates:
\begin{align}
\mathcal{L}_{\xi_{n,i}}L_{sys} &= \sum_{n'} \dot{L}_{\;\xi_{{n'},i}}^{(1)} - T_{\;\xi_{{n'},i}}^{(1)} 
\nonumber \\
&= \sum_{n'} D_{\;\xi_{n',i}}^\mu ( \dot{P}_{n',\mu}^{(1)} - Q_{n',\mu}^{(1)}  ),
\end{align}
results on the set of 3-$n$ independent equations:
\begin{align}
\Psi_{i_n}  &= \sum_{{n'} \neq n} \Big\{
 D_{\;\xi_{n',i}}^\mu \Big[ 
\Big(\square_{n'}^\alpha m_{n'} \dot{x}_{n';\alpha}\Big)g^\nu_\mu
\nonumber \\
&- \frac{1}{2}\frac{\partial m_{n'}}{\partial x_{n'}^\mu}\dot{x}^\nu_{n'}
-\frac{d}{d\tau} \Big(
\frac{\partial A^\nu}{\partial \dot{x}_{n'}^\mu}
\Big) 
- \frac{\partial A_\mu}{\partial x_{n';\nu}}
+ \frac{\partial A^\nu}{\partial x_{n'}^\mu}
 \Big]
\nonumber \\
&+D_{\;\xi_{n,i}}^\mu \Big[
\frac{d}{d\tau} \Big(
\frac{\partial A^\nu}{\partial \dot{x}_{n}^\mu}
\Big) 
+ \frac{\partial A_\mu}{\partial x_{n';\nu}}
- \frac{\partial A^\nu}{\partial x_{n}^\mu} 
  \Big]
\Big\} \dot{x}_{n';\nu}
\nonumber \\
& +\Big\{
 D_{\;\xi_{n',i}}^\mu \Big[ 
 m_{n'}g^\nu_\mu - \frac{\partial A^\nu}{\partial x_{n'}^\mu}
  \Big]
\nonumber \\
&+D_{\;\xi_{n,i}}^\mu \Big[ 
\frac{\partial A^\nu}{\partial x_{n}^\mu}
+\frac{\partial A_\mu}{\partial x_{n';\nu}}
\Big]
\Big\} \ddot{x}_{n';\nu}
\nonumber \\
&+ D_{\;\xi_{n',i}}^\mu \Big(\square_{n'}^{\dot{\alpha}} m_{n'} \ddot{x}_{n';\alpha}\Big)\dot{x}_{n';\mu}.
\label{RotConstraintEq}
\end{align}
Together with the least action principle in angular coordinates:
\begin{equation}
\mathcal{L}_{\xi_{n,i}}L_{sys} = \mathcal{L}_{\xi_{n,i}}L_{sys}( \sum_{n'} \frac{1}{2}m_{n'} \dot{x}^\nu_{n'} \dot{x}_{n';\nu} -A^\nu \dot{x}_{n';\nu} ) =0, \label{RotLeastActionEq}
\end{equation}
we 6-$n$ more independent equations to solve the problem.

We are then, in the presence of a non-holonomic constraint problem. The equations constraint the generalized coordinates of the $n$-VMVF system and they can be written as 
\begin{equation}
f(\ddot{\mathbf{r}}_1, \; \ddot{\mathbf{r}}_2 ...,\dot{\mathbf{r}}_1, \; \dot{\mathbf{r}}_2 ..., \mathbf{r}_1, \; \mathbf{r}_2 ...) = 0,
\end{equation}
showing that we are in the presence of two second order constrained Lagrangians.

\subsection{The mass and field's functions for n-VMVF systems.}
We define the derivative of the masses and the field derivative as the new unknown quantities added to the problem. However, both Lagrangians also depend on the mass of the particle $m_n$ and the vector potential $A^\nu$. Because of that, we need to express both functions with expressions depending on their derivatives. Using the Taylor's series expansion, we can write the mass $m_n(\{x^\mu\}, \{\dot{x}^\mu\})$ and the field $A^\nu(\{x^\mu\}, \{\dot{x}^\mu\})$ as:
\begin{align}
&m_n = m_n(0) + \frac{\partial m_n}{\partial x^\mu_n} \Big |_0 x_n^\mu 
+ \frac{\partial m_n}{\partial \dot{x}^\mu_n} \Big |_0 \dot{x}_n^\mu
\nonumber \\
&+ \frac{1}{2!} \frac{\partial^2 m_n}{\partial x^\mu_n \partial x^\alpha_n}\Big |_0 x_n^\mu x_n^\alpha
+ \frac{1}{2!} \frac{\partial^2 m_n}{\partial \dot{x}^\mu_n \partial \dot{x}^\alpha_n}\Big |_0 \dot{x}_n^\mu \dot{x}_n^\alpha
\nonumber \\
&+ \frac{1}{2!} \frac{\partial^2 m_n}{\partial \dot{x}^\mu_n \partial x^\alpha_n}\Big |_0 \dot{x}_n^\mu x_n^\alpha  ...
\label{MassFunctionSeries}
\end{align}
 and
\begin{align}
&A^\nu = A^\nu(0) + \sum_n \frac{\partial A^\nu}{\partial x^\mu_n} \Big |_0 x_n^\mu 
+ \frac{\partial A^\nu}{\partial \dot{x}^\mu_n} \Big |_0 \dot{x}_n^\mu 
\nonumber \\
&+ \sum_{n'} \frac{1}{2!} \frac{\partial^2 A^\nu}{\partial x^\mu_n \partial x^\alpha_{n'}}\Big |_0 x_n^\mu x_{n'}^\alpha
+ \frac{1}{2!} \frac{\partial^2 A^\nu}{\partial \dot{x}^\mu_n \partial \dot{x}^\alpha_{n'}}\Big |_0 \dot{x}_n^\mu  \dot{x}_{n'}^\alpha
\nonumber \\
&+ 2\frac{1}{2!} \frac{\partial^2 A^\nu}{\partial x^\mu_n \partial \dot{x}^\alpha_{n'}}\Big |_0 x_n^\mu \dot{x}_{n'}^\alpha ... \;. \label{FieldFunctionSeries}
\end{align}
Based on the obtained constraints, we can retain terms only up to the first derivative. The future developments and comparison with experimental data should verify the effective order of the series needed and how the present approach should be modified.

%However, the terms with higher powers included on the Lagrangian must be expressed as functions of the already defined degree of freedom, $e.i.$ the particle coordinates and the derivatives of the mass and field with coordinates, or theirs derivative with time. For example, the second order derivative of mass with the coordinate can be isolated from the set of equations:
%\begin{align}
%\frac{d}{d \tau} \Big( \frac{\partial m}{\partial x^\mu_n}\Big) 
%&= \frac{\partial^2 m}{\partial x^\mu_n \partial x^\nu_n} \dot{x}^\nu_n
%+ \frac{\partial^2 m}{\partial x^\mu_n \partial \dot{x}^\nu_n} \ddot{x}^\nu_n
%\nonumber \\
%\frac{d}{d \tau} \Big( \frac{\partial m}{\partial \dot{x}^\mu_n}\Big) 
%&= \frac{\partial^2 m}{\partial \dot{x}^\mu_n \partial x^\nu_n} \dot{x}^\nu_n
%+ \frac{\partial^2 m}{\partial \dot{x}^\mu_n \partial \dot{x}^\nu_n} \ddot{x}^\nu_n
%\end{align}

\subsection{The mass and fields derivatives for n-VMVF systems.} \label{massFieldApprox}

To obtain the inertial and applied forces acting over each particle, we set some constraints to the new degrees of freedom added to the problem
\begin{align}
\{\frac{\partial m_n}{\partial x_n^\mu }\}, \quad
\{\frac{\partial m_n}{\partial \dot{x}_n^\mu }\} , \quad
\{\frac{\partial A^\nu}{\partial x_n^\mu }\},\quad
\{\frac{\partial A^\nu}{\partial \dot{x}_n^\mu }\}. \label{newDegreeFreedom}
\end{align}
For example, we assumed that the mass of the particle depends only on the position and velocity of the own particle. From the analytic point of view, the only requirement these new quantities must have is that the system of equations must be solvable. It means that the total number of variables must match to the total number of equations. Now that we have the total number of equations, we can propose a more accurate dependency for derivatives \ref{newDegreeFreedom}.

We have a total of $14n$ equations: the $4n$ and $3n$ extended Lagrange equations for the Lorentzian $x_n^\nu$ and angular coordinates $\xi_n$, plus the $4n$ and $3n$ constraint equations in both coordinate systems, respectively. Nevertheless, 2$n$ equations are restricted to the Lorentz constraints, letting the total number of equations equal to $12n$. On the other hand, if we sum all possibilities in \ref{newDegreeFreedom}, we will have a total of $40n$ independent new variables.

The $4n$ coordinates of the particle $x_n^\nu$ are constrained by the $n$ Lorentz relation leaving $3n$ independent variables. So to solve the system of equation, we need then to attribute the left 9$n$ degrees of freedom into the mass and field derivatives. This characteristic reflects the constructive character of this methodology. We propose to divide the $9n$ number of variables as:
\begin{enumerate}
\item $3n $ for $ \{ \frac{\partial m}{\partial x_n^\nu}\}  $ and for $ \{ \frac{\partial m}{\partial \dot{x}_n^\nu}\}  $
\item $3n $ for $ \{ \frac{\partial A^\mu}{\partial \dot{x}_n^\nu}\}  $
\item $3n $ for $ \{ \frac{\partial A^\mu}{\partial x_n^\nu}\}  $.
\end{enumerate}
We analyze different approaches according to this division. Other distributions of the number of variables and others approximations can be applied as long the number of independent variables remains equal to the number of equations; however, the results and their physical interpretation will be according to that choice

\begin{itemize}
\item Masses derivatives $\frac{\partial m_n}{x^\nu_{n'}} $ and $\frac{\partial m_n}{\dot{x}^\nu_{n'}} $.

From the beginning of this work, we restricted the variation of the mass of particles depending only on its own particle's position which means
\begin{equation*}
\frac{\partial m_n}{x^i_{n'}} \equiv \frac{\partial m_n}{x^i_{n'}}\delta_{nn'} \qquad \forall x_i = \{x,y,z\} \neq 0. 
\end{equation*}
These derivatives sum $3n$ independent variables to the system and they correspond to the structural variation of the mass. The relativistic theory was included in our approach adding the variations of the mass related to the inertial frame. As discussed, such variations depend on the velocity of the own particle only, which means that
\begin{equation}
\frac{\partial m_n}{\dot{x}^\nu_{n'}} \equiv \frac{\partial m_n}{\dot{x}^\nu_{n'}}\delta_{nn'} \neq 0\;\;\; \text{and} \;\;\; \frac{\partial m_n}{\dot{x}^0_{n'}} =0.
\end{equation} 
As showed in the previous section, the $4n$ derivatives of mass with velocity are constrained to the $4n$ equations \ref{mass0compConstraint}:
\begin{equation*}
\frac{\partial m}{\partial \dot{x}^\nu} \ddot{x}^\nu \dot{x}_\mu + m\ddot{x}_\mu =0,
\end{equation*}
which let unchanged the number of independent variables to $3n$, remaining 6$n$ more to define. 

\item Field derivatives with the velocity $\frac{\partial A^\nu}{\partial \dot{x}^\mu_{n}}$.

Every component of the vector field depends on the positions and the velocities of all particles. In that case, we have $4\times 4n = 16n$ variables related to the derivative with velocity. For clarity purposes, let us explicitly separate the variables $\frac{\partial A^\nu}{\partial \dot{x}^\mu_{n}}$ as
\begin{align}
\Big \{\frac{\partial A^\nu}{\partial \dot{x}^\mu_{n}} \Big \} &= 
\Big \{\frac{\partial A^0}{\partial \dot{x}^0_{n}} \Big \}
+ \Big \{\frac{\partial A^i}{\partial \dot{x}^0_{n}} \Big \}
+ \Big \{\frac{\partial A^0}{\partial \dot{x}^i_{n}} \Big \}
+ \Big \{\frac{\partial A^i}{\partial \dot{x}^j_{n}} \Big \} 
%\quad \forall \; i,j =1,2,3 
\nonumber \\
\{16n\} &= \;\;\; \{n\} \quad +\;\;  \{3n\} \; + \;\; \{3n\} \;\;\; + \;\;\{9n\}.
 \label{fieldVarDiv}
\end{align}

As discussed in section \ref{classicalAssump1} and resumed in equations \ref{relConstraint1}, we suppose that the component of the field $A^0$ has the same behavior as the scalar field $\phi(\mathbf{x})$ on the Electromagnetic field, which is related to the field's fixed sources contribution. We have then 
\begin{equation*}
\frac{\partial A^0}{\partial \dot{x}^\nu_{n}} = 0.
\end{equation*}
This condition removes the first and the third set from the independent group of variables in \ref{fieldVarDiv} letting the total variables number equal to $12n$. On the other hand, the derivative $\frac{\partial A^\nu}{\partial \dot{x}^\mu_{n}} $ may be inferred by also analyzing the Electromagnetic field were the component of the field depends on the same component of the velocity of the particle. We extrapolate this dependency to our universal field $e.i$
\begin{equation}
\frac{\partial A^\nu}{\partial \dot{x}^\mu_{n}} \equiv \frac{\partial A^\nu}{\partial \dot{x}^\mu_{n}} \delta^\mu_\nu.\label{fieldVelDep}
\end{equation}
This approach set eliminates the second group of variables in \ref{fieldVarDiv} and decreases the number of variables of the last group to $3n$ as needed.

\item Field derivative with the position $\frac{\partial A^\nu}{\partial x^\mu_{n}}$.

We have $4\times 4n = 16n$ variables related to the derivative with the position. Given the different natures and forms for the field defined along the development of physics, this is the more complex degree of freedom, and because of that, various approaches can be made for different physical systems. The propositions can also depend on the numbers of particles of the system. The field derivative with $x_n^0$ is chosen to be constrained by the Gauge invariance, as shown in equation \ref{relConstraint1} and discussed in section \ref{classicalAssump1}:
\begin{equation*}
\partial_{n,\nu} A^\nu \equiv \sum_\nu \frac{\partial A^\nu}{\partial x_n^\nu} = 0.
\end{equation*}
Initially, for the position derivative, we have $4\times 4n = 16n$ independent variables, explicitly we can divided them as
\begin{align}
\Big \{\frac{\partial A^\nu}{\partial x^\mu_{n}} \Big \} &= 
\Big \{\frac{\partial A^0}{\partial x^0_{n}} \Big \}
+ \Big \{\frac{\partial A^i}{\partial x^0_{n}} \Big \}
+ \Big \{\frac{\partial A^0}{\partial x^i_{n}} \Big \}
+ \Big \{\frac{\partial A^i}{\partial x^j_{n}} \Big \} 
%\quad \forall \; i,j =1,2,3
\nonumber \\
\{16n\} &= \;\;\; \{n\} \quad +\;\;  \{3n\} \; + \;\; \{3n\} \;\;\; + \;\;\{9n\}
 \label{fieldVarDiv1}
\end{align}
We find essential to discuss 3 different cases:
\begin{enumerate}
\item The derivative of the field depending on the same direction of the vector of the position of the particle.

Just as the approach for the derivative of the field with respect to the velocity, we can also set derivative of the field with the particle's position dependent on the same component than the position of the particle as:
 \begin{equation}
\frac{\partial A^\nu}{\partial x^\mu_{n}} \equiv \frac{\partial A^\nu}{\partial x^\mu_{n}} \delta^\mu_\nu.
\end{equation}

By using this approach, the second and third set of variables in \ref{fieldVarDiv1} are set to zero, while the number of the fourth group of variables is reduced to $3n$ for a total of $4n$ independent. Finally, taking into account the Gauge condition \ref{relConstraint1} equations, we have defined 3-$n$ more independent variables, completing the $12n$ variables of the system described with $12n$ equations.

This approach may have a weak physical meaning, but it can be used to study isolated $n$-VMVF systems with any number of particles. Indeed, the field will not have the ordinary physical meaning depending on distances between particles, but it shall play its role as the mathematical entity that ``connect'' particles and ``transport information''  between them, whose only restriction is that system of equations is solvable.

\item Field component derivative depending on the distance between particles. 

The most accepted idea on physics for any field connecting particles is that it depends on the distance between particles. This type of fields is known in the literature as central fields. In this case, the degree of freedom of the system is $\frac{\partial A^\nu}{\partial s_{ij}}$ where the Lorentz invariant $s_{ij}$ is the distance between particle $i$ and $j$, defined as:
\begin{equation}
s_{ij}=\sqrt{\sum_\nu(x^\nu_i-x^\nu_j)^2}.
\end{equation}

The total number of the derivatives of the field with distances needs to be equal to 3$n$. The number of distances is the number of 2-combinations of $n$ particles:
\begin{equation}
N(s_{ij}) = \binom{n}{2} = \frac{n(n-1)}{2}
\end{equation}
Under this approach, if the number of particles increases enough, the number of the derivatives of field with the distance between particle will surpass the 3$n$ variables limit at some point. The Gauge $n$-conditions now have the form:
\begin{equation}
\partial_{n,\nu} A^\nu 
\equiv \sum_\nu \frac{\partial A^\nu}{\partial x_n^\nu} = 
\sum_{n'\neq n} \frac{\partial A^\nu}{s_{nn'}} \frac{s_{nn'}} {\partial x_n^\nu}
=0.
\end{equation}
Note that the gauge conditions are independent for each particle except for the number of particles equal 2.

We study two different cases:
\begin{enumerate}
\item The most general case is when all derivatives $\frac{\partial A^\nu}{\partial s_{ij}}$ are different: 
\begin{equation}
\frac{\partial A^0}{\partial s_{ij}} 
\neq \frac{\partial A^1}{\partial s_{ij}} 
\neq \frac{\partial A^2}{\partial s_{ij}}
\neq \frac{\partial A^3}{\partial s_{ij}}.
\end{equation}
Using the distance between particles, the number of independent variables is the number of components of the field $(4)$ times the number of distances $n$ minus $n$ corresponding to the number of Gauge conditions for every particle. If we equate this number of variables to our limit $3n$ we have:
\begin{equation}
4 N(s_{ij}) -n = 4 \frac{n(n-1)}{2} -n = 3n \qquad  \therefore n= 3.
\end{equation}
The result means that we can successfully describe an $3$-VMF systems assuming field depending on the distances between particles. If the number of particles is greater than 3, then central field approach cannot be used. 

If the number of particles is $N=2$, it is possible to describe the field using the $s_{ij}$'s dependency; however, this time the number of equations is greater than the number of variables, and because of that, others degree of freedom must be added. In the case of two particles, the number of distance is
$N(s_{ij}) = n(n-1)/ 2 = 1$ and the gauge $2$-conditions reduce to 1 since they are not independent:
\begin{equation}
0=\partial_{1,\nu} A^\nu
=\frac{\partial A^\nu}{s_{12}} \frac{s_{12}} {\partial x_1^\nu}
=- \frac{\partial A^\nu}{s_{12}} \frac{s_{12}} {\partial x_2^\nu} 
= -\partial_{2,\nu} A^\nu.
\end{equation}
The number of independent variables is $4 N(s_{ij}) -n_{Gauge} = 3$. Now we are short on the number of independent variables. We must then, add another dependency. We can modify, for example, the dependency of the field with velocity and instead of equation \ref{fieldVelDep}, we propose
\begin{equation}
\frac{\partial A^\nu}{\partial \dot{x}^\mu_{n}} = \frac{\partial A^i}{\partial \dot{x}^j_{n}} \delta^i_j \quad \text{where }\; i,j=1,2,3.\label{fieldVelDep1}
\end{equation}
Now, the second group of variables of expression \ref{fieldVarDiv} 
$ \Big \{\frac{\partial A^i}{\partial \dot{x}^0_{n}} \Big \}$
is no longer zero and add $6$ more variables. In that case, we can set the constraint
\begin{equation}
\frac{\partial A^i}{\partial \dot{x}^0_{1}} = \frac{\partial A^i}{\partial \dot{x}^0_{2}}
\end{equation}
and reduce them to 3 to obtain the finals 6 independent variables for the system.

\item Other case is when the three degrees of freedom related to the field derivative component $\frac{\partial A^\nu}{\partial s_{ij}}$ have the same dependency, $e.i$
\begin{equation}
\frac{\partial A^0}{\partial s_{ij}} 
= \frac{\partial A^1}{\partial s_{ij}} 
= \frac{\partial A^2}{\partial s_{ij}}
= \frac{\partial A^3}{\partial s_{ij}}.
\end{equation}
This approach resembles Electromagnetic vector field where all its component depends on distance as $s_{ij}^{-2}$. In this case, the number of distances cannot be greater that $3n$
\begin{equation}
\frac{n(n-1)}{2} - n = 3n \quad n(n-9) \therefore n= 9.
\end{equation}
We can, then, describe $n$-VMVF system with these conditions and previous restrictions, up to 9 particles. In the case of a lesser number of particles, others mass or field derivative need to be added as variables of the system, same as the previous example.
\end{enumerate}
\end{enumerate}
\end{itemize}

The primary assumption of this work is to consider the mass of the particle and the field as unknown variables, being the conservation of linear and angular momentums and the principle of least action their only restrictions. However, we can not avoid assuming some forms for the derivatives of those functions with particle position and velocities, so the system remains solvable. Nevertheless, the initial assumption has lost generality, the number of degrees of freedom related to the mass of the field is higher than if a fixed form is chosen.

\section{The second order constrained Lagrangian.}
On the last sections, we obtain 2 set of equations of motions for $n$-VMVF systems; each one for every independent set of coordinates as shown on eqs. \ref{TransConstraintEq}, \ref{TransLeastActionEq}, \ref{RotConstraintEq} and \ref{RotLeastActionEq}. The constraints from equations \ref{RotConstraintEq} and \ref{TransConstraintEq} have a second order dependency like
\begin{align}
\varphi_i(x,y_1,y_2,...y_n,\dot{y}_1,\dot{y}_2,...\dot{y}_n,\ddot{y}_1,\ddot{y}_2,...\ddot{y}_n)=0 
\nonumber \\
(i=1,2,....m;m<n),
\end{align}
then, the integrand must also have the same dependency as
\begin{equation}
L_{sys}(x,y_1,y_2,...y_n,\dot{y}_1,\dot{y}_2,...\dot{y}_n,\ddot{y}_1,\ddot{y}_2,...\ddot{y}_n)=0.
\end{equation}
This dependency implies the necessity to expand the equations of Euler-Lagrange up to the second order. 

It is not difficult to obtain the extended Euler-Lagrange equations up to the second order, as shown in the textbook of R. Courant and D. Hilbert ``Methods of Mathematical Physics'' \cite{Courant53physics}:
\begin{equation}
\frac{d^2}{dx^2}\Big( \frac{\partial F}{\partial \ddot{y}_i}\Big) - \frac{d}{dx}\Big( \frac{\partial F}{\partial \dot{y}_i}\Big) + \frac{\partial F}{\partial y_i} =0\label{extLagrangeEq2Order}
\end{equation}
In general can be proved that for higher $n$-order dependency of the variational 
\begin{equation}
\int F(x, y,y^{(1)},... y^{(n)})dx,
\end{equation}
the Euler-Lagrange equations have the form:
\begin{equation}
\sum_{i=1}^n (-1)^n \frac{d^n}{dx^n}\Big( \frac{\partial F}{\partial y_i^{(n)}}\Big) + \frac{\partial F}{\partial y_i} = 0.
\end{equation}

The existence of constraints like  Eq \ref{TransConstraintEq} and \ref{RotConstraintEq} means that some virtual displacements of the generalized coordinates from the second-order Lagrangian
\begin{equation}
L(x, y_1,y_2...y_n,\dot{y}_1,\dot{y}_2...\dot{y}_n,\ddot{y}_1,\ddot{y}_2...\ddot{y}_n) 
\end{equation}
being
\begin{equation}
\{y_{n}\} \equiv \{\mathbf{r}_n,\frac{\partial m_n}{\partial x_{n,i}},\frac{\partial \mathbf{A}}{\partial x_{n,i}}, \frac{\partial \mathbf{A}}{\partial \dot{x}_{n,i}}\},
\end{equation}
are not independent. In this case, the fundamental lemma of calculus of variations to obtain Euler-Lagrange equations can no longer be applied, and the extended Lagrange equations \ref{extLagrangeEq2Order} are no longer valid.

\subsection{The Lagrange method of the undetermined multipliers up to the second order.}

One of the most efficient procedures to treat these displacements is the well-known method of the undetermined multipliers \citep{elsgoltz1977} developed by Lagrange. This method is a strategy to solve the extremal problem for functions or functionals subject to equality constraint relations. Thus, given the functional
\begin{equation}
J = \int_{x_0}^{x_1} F(x, y_1,y_2...y_n,\dot{y}_1,\dot{y}_2...\dot{y}_n,\ddot{y}_1,\ddot{y}_2...\ddot{y}_n)dx,
\end{equation}
and the equality constraint relations
\begin{align}
\varphi_i(x, y_1,y_2...y_n,\dot{y}_1,\dot{y}_2...\dot{y}_n,\ddot{y}_1,\ddot{y}_2...\ddot{y}_n)=0
\nonumber \\
(i=1,2,....m;m<n),
\end{align}
under a proper selection of constants $\lambda_i$, it is proved that if functions $y_j\;\; j=(1,2,..n)$ are extremes of the problem then, they also are extremes for the generalized functional:
\begin{equation}
J^* = \int_{x_0}^{x_1} \Big( F + \sum_{i=1}^m \lambda_i(x)\varphi_i \Big)dx = \int_{x_0}^{x_1} F^*dx.
\end{equation}
The system is then solved considering functions $y_1$, $y_2$..., $y_n$ and $\lambda_1$, $\lambda_2$ ..., $\lambda_m$ as arguments of variational $J^*$.

The proof is similar to the method exposed on reference \citep{elsgoltz1977} for the $\varphi (y_i,\dot{y}_i)$ dependence. The constraint equations $\varphi_i$ are independent, so they satisfy
\begin{equation}
\frac{D(\varphi_1, \varphi_2,....\varphi_m) }{D( \ddot{y}_1,\;\ddot{y}_2\;....\ddot{y}_m)}\neq 0.
\end{equation}
In that case, $\ddot{y}_1$, $\ddot{y}_2$ ... $\ddot{y}_m$ variables can be determined as
\begin{align}
\ddot{y}_i = \Psi_i (x, y_1,y_2...y_n,\dot{y}_1,\dot{y}_2...\dot{y}_n,\ddot{y}_1,\ddot{y}_2...\ddot{y}_n) 
\nonumber \\
 i = 1,2...,m,
\end{align}
being $y_i \;(i=m+1,m+2...n)$ functions whose variations $\delta y_i$ are arbitrary and, also with the variations of its derivatives $\delta \dot{y}_i$ and $\delta \ddot{y}_i$, vanish at $x_0$ and $x_1$ points. If $y_1...y_n$ are arbitrary functions that satisfy the $m$ equations $\varphi_i$, then the variation of these constraints is
\begin{equation}
\delta \varphi_i=\sum_{j=1}^n \frac{\partial \varphi_i}{\partial y_j} \delta y_j + \frac{\partial \varphi_i}{\partial \dot{y}_j} \delta \dot{y}_j + \frac{\partial \varphi_i}{\partial \ddot{y}_j} \delta \ddot{y}_j =0,
\end{equation}
where the higher order of the variations $\delta y_j $, $\delta \dot{y}_j$ and $\delta \ddot{y}_j$ are omitted because their influence are minimized when computing the variation of the variational function where only the first order of $\delta y_j $, $\delta \dot{y}_j$ and $\delta \ddot{y}_j$ matters \citep{elsgoltz1977}.  

Multiplying by the undetermined factor $\lambda_i(x)$ and integrating over $dx$:
\begin{align}
&\int_{x_0}^{x_1} \lambda_i(x)\delta \varphi_i dx = 0
\nonumber \\
&= \int_{x_0}^{x_1} \Big\{ \sum_{j=1}^n \lambda_i(x)\frac{\partial \varphi_i}{\partial y_j} \delta y_j + \lambda_i(x)\frac{\partial \varphi_i}{\partial \dot{y}_j} \delta \dot{y}_j 
\nonumber \\
&\qquad \qquad + \lambda_i(x)\frac{\partial \varphi_i}{\partial \ddot{y}_j} \delta \ddot{y}_j\Big\}dx.
\end{align}
Integrating by parts, the second integrand has the form
\begin{align}
\int_{x_0}^{x_1} \sum_{j=1}^n \lambda_i(x)&\frac{\partial \varphi_i}{\partial \dot{y}_j} \delta \dot{y}_j dx
=  \sum_{j=1}^n \lambda_i(x)\frac{\partial \varphi_i}{\partial \dot{y}_j} \delta {y}_j\Big\vert_{x_0}^{x_1} 
\nonumber \\
&\quad- \int_{x_0}^{x_1} \sum_{j=1}^n \frac{d}{dx}\Big( \lambda_i(x)\frac{\partial \varphi_i}{\partial \dot{y}_j}\Big) \delta y_j dx \nonumber \\
&= - \int_{x_0}^{x_1} \sum_{j=1}^n \frac{d}{dx}\Big( \lambda_i(x)\frac{\partial \varphi_i}{\partial \dot{y}_j}\Big) \delta y_j dx
\end{align}
and the third integrand results as
\begin{align}
&\int_{x_0}^{x_1} \sum_{j=1}^n \lambda_i(x)\frac{\partial \varphi_i}{\partial \ddot{y}_j} \delta \ddot{y}_j dx 
=  \sum_{j=1}^n \lambda_i(x)\frac{\partial \varphi_i}{\partial \ddot{y}_j} \delta \dot{y}_j\Big\vert_{x_0}^{x_1} 
\nonumber \\
&\quad- \int_{x_0}^{x_1} \sum_{j=1}^n \frac{d}{dx}\Big( \lambda_i(x)\frac{\partial \varphi_i}{\partial \ddot{y}_j}\Big) \delta \dot{y}_j dx 
\nonumber \\
&= \sum_{j=1}^n -\frac{d}{dx}\Big( \lambda_i(x)\frac{\partial \varphi_i}{\partial \ddot{y}_j}\Big) \delta y_j\Big\vert_{x_0}^{x_1} 
\nonumber \\
& \qquad + \int_{x_0}^{x_1} \sum_{j=1}^n \frac{d^2}{dx^2}\Big( \lambda_i(x)\frac{\partial \varphi_i}{\partial \ddot{y}_j}\Big) \delta y_j dx 
\nonumber \\
&= \int_{x_0}^{x_1} \sum_{j=1}^n \frac{d^2}{dx^2}\Big( \lambda_i(x)\frac{\partial \varphi_i}{\partial \ddot{y}_j}\Big) \delta y_j dx,
\end{align}
where the variations $\delta y_i$ and its derivatives $\dot{y}_i$ are zero at points $x_0$ and $x_1$. Putting all together, we have:
\begin{align}
\int_{x_0}^{x_1}& \lambda_i(x)\delta \varphi_i dx = 
\int_{x_0}^{x_1} \sum_{j=1}^n \Big[ \lambda_i(x)\frac{\partial \varphi_i}{\partial y_j} 
\nonumber \\
&- \frac{d}{dx}\Big( \lambda_i(x)\frac{\partial \varphi_i}{\partial \dot{y}_j}\Big)  
+ \frac{d^2}{dx^2}\Big( \lambda_i(x)\frac{\partial \varphi_i}{\partial \ddot{y}_j}\Big) \Big] \delta y_j dx.
\end{align}
Adding these $m$ equations to the $\delta J$ variation
\begin{equation}
\delta J = \int_{x_0}^{x_1}  \sum_{j=1}^n \Big[ \frac{d^2}{dx^2}\Big(\frac{\partial F}{\partial \ddot{y}_j} \Big) - \frac{d}{dx}\Big(\frac{\partial F}{\partial \dot{y}_j} \Big) + \frac{\partial F}{\partial y_j}\Big]  \delta y_j dx 
\end{equation}
we obtain
\begin{equation}
\delta J = \int_{x_0}^{x_1}  \sum_{j=1}^n \Big[ \frac{d^2}{dx^2}\Big(\frac{\partial F^*}{\partial \ddot{y}_j} \Big) - \frac{d}{dx}\Big(\frac{\partial F^*}{\partial \dot{y}_j} \Big) + \frac{\partial F^*}{\partial y_j}\Big]  \delta y_j dx \label{EulerLagrangeEq1}
\end{equation}
being
\begin{equation}
F^*= F+\sum_{i=1}^m \lambda_i(x)\varphi_i.
\end{equation}
The variations $\delta y_i$ are not arbitrary since they are restricted to constraints $\varphi_i$. However, factors $\lambda_i(x)$ can be chosen to satisfy
\begin{equation}
\frac{d^2}{dx^2}\Big(\frac{\partial F^*}{\partial \ddot{y}_j} \Big) - \frac{d}{dx}\Big(\frac{\partial F^*}{\partial \dot{y}_j} \Big) + \frac{\partial F^*}{\partial y_j}=0 \;\;\; (j=1,2,...m),
\end{equation}
defining a set of linear equations depending on
\begin{equation}
\lambda_i,\;\;\;\; \frac{\partial \lambda_i}{\partial x} \;\; \text{and}\;\; \frac{\partial^2 \lambda_i}{\partial x^2}.
\end{equation}
If $\delta y_j$, $(j=1,2,...m)$, are chosen, without any loss of generality, as the nonarbitrary variations then equations \ref{EulerLagrangeEq1} reduce to
\begin{equation}
\delta J = \int_{x_0}^{x_1}  \sum_{j=m+1}^n \Big[ \frac{d^2}{dx^2}\Big(\frac{\partial F^*}{\partial \ddot{y}_j} \Big) - \frac{d}{dx}\Big(\frac{\partial F^*}{\partial \dot{y}_j} \Big) + \frac{\partial F^*}{\partial y_j}\Big]  \delta y_j dx, \label{IndepVariations}
\end{equation}
where the $\delta y_j$ $(j=m+1,m+2,...n)$ are now independent, which allows the application of the fundamental lemma of the calculus of variations and obtain:
\begin{align}
\frac{d^2}{dx^2}\Big(\frac{\partial F^*}{\partial \ddot{y}_j} \Big) - \frac{d}{dx}\Big(\frac{\partial F^*}{\partial \dot{y}_j} \Big) + \frac{\partial F^*}{\partial y_j}=0 
\nonumber \\
(j=m+1,m+2,...n).
\end{align}
Thereby, the functions $y_1(x), y_2(x)...y_n(x)$ that extreme the variational$J(y_1,y_2,...y_n)$ and constants $\lambda_1(x), \lambda_2(x)...\lambda_m(x)$ must satisfy the set of $n+m$ equations
\begin{equation}
\frac{d^2}{dx^2}\Big(\frac{\partial F^*}{\partial \ddot{y}_j} \Big) - \frac{d}{dx}\Big(\frac{\partial F^*}{\partial \dot{y}_j} \Big) + \frac{\partial F^*}{\partial y_j}=0 \;\;\; (j=1,2,...n)
\end{equation}
and
\begin{align}
\varphi_i(x, y_1,y_2...y_n,\dot{y}_1,\dot{y}_2...\dot{y}_n,\ddot{y}_1,\ddot{y}_2...\ddot{y}_n)=0
\nonumber \\
 (i=1,2,....m).
\end{align}

\section{The Lagrangians for the n-VMVF systems}
The Lagrange method of the undetermined multipliers has been extended to include second order constraints. We are in a position now to apply all the previous results to the $n$-VMVF isolated systems and construct the finals Lagrangians.

The second order Lagrange operators in the Lorentzian and angular coordinates are:
\begin{equation}
\mathcal{L}_{n,\nu}  \equiv \Big[ -  \frac{d^2}{d\tau^2}\frac{\partial}{\partial \ddot{x}^\nu_n}  + \frac{d}{d\tau}\frac{\partial}{\partial \dot{x}^\nu_n} - \frac{\partial}{\partial x_n^\nu}\Big],  \label{lagrangetranslationExt}
\end{equation}
and
\begin{equation}
\mathcal{L}_{n,\xi}  \equiv \Big[ - \frac{d^2}{d\tau^2}\frac{\partial}{\partial \ddot{\xi}_n} + \frac{d}{d\tau}\frac{\partial}{\partial \dot{\xi}_n} - \frac{\partial}{\partial \xi_n}\Big], \;\; \xi=\{ \theta,\phi,\chi\}. \label{lagrangeRotationExt}
\end{equation}

Using the relations \ref{relationsMatrix}, the operator of Lagrange expressed in angular coordinates of eq \ref{lagrangeRotationExt}, takes the form:
\begin{align}
&\mathcal{L}_{n,\xi_i} =  - \frac{d^2}{d\tau^2} \Big[  D_{\;\xi_{n,i}}^\mu \square_{n,\ddot{\mu}} \Big] 
+ \frac{d}{d\tau} \Big[  D_{\;\xi_{n,i}}^\mu \square_{n,\dot{\mu}}  
\nonumber \\
&+  2\frac{d}{d\tau}\Big(  D_{\;\xi_{n,i}}^\mu \Big) \square_{n,\ddot{\mu}} \Big]
- \Big[ D_{\;\xi_{n,i}}^\mu \square_{n,\mu}  
\nonumber \\
&+  \frac{d}{d\tau}\Big(  D_{\;\xi_{n,i}}^\mu \Big)\square_{n,\dot{\mu}}  
+ \frac{d^2}{d\tau^2}\Big(  D_{\;\xi_{n,i}}^\mu \Big)\square_{n,\ddot{\mu}} \Big]
%\nonumber \\
%&= - \frac{d^2}{d\tau^2} \Big(  D_{\;\xi_{n,i}}^\mu \Big) \square_{n,\ddot{\mu}} 
%- 2\frac{d}{d\tau} \Big(  D_{\;\xi_{n,i}}^\mu \Big) \frac{d}{d\tau} \Big( \square_{n,\ddot{\mu}} \Big) 
%\nonumber \\
%&- D_{\;\xi_{n,i}}^\mu  \frac{d^2}{d\tau^2} \Big(\square_{n,\ddot{\mu}} \Big)
%+ \frac{d}{d\tau} \Big(  D_{\;\xi_{n,i}}^\mu \Big) \square_{n,\dot{\mu}}  
%\nonumber \\
%&+  D_{\;\xi_{n,i}}^\mu  \frac{d}{d\tau} \Big( \square_{n,\dot{\mu}} \Big) 
%+  2\frac{d^2}{d\tau^2}\Big(  D_{\;\xi_{n,i}}^\mu \Big) \square_{n,\ddot{\mu}} 
%\nonumber \\
%&+  2\frac{d}{d\tau}\Big(  D_{\;\xi_{n,i}}^\mu \Big) \frac{d}{d\tau} \Big( \square_{n,\ddot{\mu}} \Big) 
%-  D_{\;\xi_{n,i}}^\mu \square_{n,\mu}  
%\nonumber \\
%&-  \frac{d}{d\tau}\Big(  D_{\;\xi_{n,i}}^\mu \Big)\square_{n,\dot{\mu}}
%- \frac{d^2}{d\tau^2}\Big(  D_{\;\xi_{n,i}}^\mu \Big)\square_{n,\ddot{\mu}}.
\nonumber \\
& = - D_{\;\xi_{n,i}}^\mu  \frac{d^2}{d\tau^2} \Big(\square_{n,\ddot{\mu}} \Big) 
+  D_{\;\xi_{n,i}}^\mu  \frac{d}{d\tau} \Big( \square_{n,\dot{\mu}} \Big) 
\nonumber \\
&\quad -  D_{\;\xi_{n,i}}^\mu \square_{n,\mu}.
\end{align}
Finally,
\begin{equation}
\mathcal{L}_{n,\xi_i} = D_{\;\xi_{n,i}}^\mu \Big[ -\frac{d^2}{d\tau^2} \Big(\square_{n,\ddot{\mu}} \Big) +  \frac{d}{d\tau} \Big( \square_{n,\dot{\mu}} \Big) - \square_{n,\mu} \Big]. \label{lagrangeRotationExt1}
\end{equation}

We can apply the undetermined multipliers method of Lagrange for the rearranging the obtained equations of motions for $n$-VMVF systems, resumed in equations \ref{TransLeastActionEq},\ref{RotLeastActionEq}, \ref{TransConstraintEq} and \ref{RotConstraintEq}. According to the previous section, the solutions that minimize the former Lagrangian and satisfy the two independent set of constraint equations for each set of coordinates, also minimize the Lagrangians:
\begin{equation}
L_T = L_{sys} + \sum_{n} \lambda^\nu_n \Phi_{\nu_n},
\quad
L_R = L_{sys} + \sum_{i,n} \beta_{i_n} \Psi_{i_n},
\end{equation}
where the constructed Lagrangian with the Lorentzian coordinates is represented by $L_T$ and the Lagrangian constructed using the angular as $L_R$. The $T$ and $R$'s index stand for the translation and rotation transformations as the operation described with the Lorentzian and the angular coordinates, respectively. 

The relativistic equations \ref{relatRelation}, \ref{relConstraint1} and \ref{mass0compConstraint} related to the Lorentz and the Gauge relations and the conservation of the relativistic momentum for a single particle
\begin{align}
\Omega_n &\equiv x^\nu_n x_{\nu;n} -R^2_n =0 ,
\nonumber \\
\Upsilon_n &\equiv \partial_{n;\nu} A^\nu=0
\nonumber \\
\Theta_{n;\mu} &\equiv \frac{\partial m_n}{\partial \dot{x}^\nu_n} \ddot{x}^\nu_n \dot{x}_{n;\mu} + m\ddot{x}_{n;\mu} =0, \label{relConstraintFinal}
\end{align}
also restrict the motion of the system, and they should also be included in the Lagrangians. Those constraints are present on both sets of Lagrangian equations. In that case, is not hard to show that they should be added to each Lagrangian using the same coefficients like
\begin{align}
L_T &= L_{sys} + \sum_{n} \lambda^\nu_n \Phi_{\nu_n} + \omega_n \Omega_n + \epsilon_n \Upsilon_n + \eta^\mu_n \Theta_{n;\mu}
\\
L_R &= L_{sys} + \sum_{i,n} \beta_{i_n} \Psi_{i_n}.
\end{align}
or 
\begin{align}
L_T &= L_{Rel} + \sum_{n} \lambda^\nu_n \Phi_{\nu_n}
\\
L_R &= L_{Rel} + \sum_{i,n} \beta_{i_n} \Psi_{i_n},
\end{align}
where the relativistic constraints are included on a more general Lagrangian 
\begin{equation*}
L_{Rel} =  L_{sys} + \sum_{n} \omega_n \Omega_n + \epsilon_n \Upsilon_n + \eta^\mu_n \Theta_{n;\mu}.
\end{equation*}
%\begin{widetext}
If we replace the mass and field function for theirs the series expansion for the mass and the field function from equations \ref{FieldFunctionSeries} and \ref{MassFunctionSeries} up to linear terms, the constraint equations \ref{TransConstraintEq} and \ref{RotConstraintEq} now have the form
\begin{align}
\Phi_{\mu_n} &= \sum_{{n'} \neq n}  \Big[
\Big(\square_{n'}^\alpha m_{n'} \dot{x}_{n';\alpha}\Big)g^\nu_\mu
- \frac{1}{2}\frac{\partial m_{n'}}{\partial x_{n'}^\mu}\dot{x}^\nu_{n'}
+ \frac{d}{d\tau} \Big(\frac{\partial A^\nu}{\partial \dot{x}_{n}^\mu}
- \frac{\partial A^\nu}{\partial \dot{x}_{n'}^\mu}
\Big) 
\nonumber \\
&+ \frac{\partial A^\nu}{\partial x_{n'}^\mu}
- \frac{\partial A^\nu}{\partial x_{n}^\mu}
\Big]\dot{x}_{n';\nu}
+ \Big[ 
\Big(m_{n'}(0) + \frac{\partial m_{n'}}{\partial x^\alpha_{n'}} {x}_{n'}^\alpha
+ \frac{\partial m_{n'}}{\partial \dot{x}^\alpha_{n'}}  \dot{x}_{n'}^\alpha \Big) g^\nu_\mu 
\nonumber \\
&+ \frac{\partial A^\nu}{\partial x_{n}^\mu}
- \frac{\partial A^\nu}{\partial x_{n'}^\mu}
+ \frac{\partial A_\mu}{\partial x_{n'}^\nu}
\Big] \ddot{x}_{n';\nu}
+ \square_{n'}^{\dot{\mu}} m_{n'} \ddot{x}_{n';\mu}  \dot{x}_{n';\nu}
\label{TransConstraintEqApprx}
\end{align}
and
\begin{align}
\Psi_{i_n}  &= \sum_{{n'} \neq n} \Big\{
 D_{\;\xi_{n',i}}^\mu \Big[ 
\Big(\square_{n'}^\alpha m_{n'} \dot{x}_{n';\alpha}\Big)g^\nu_\mu
- \frac{1}{2}\frac{\partial m_{n'}}{\partial x_{n'}^\mu}\dot{x}^\nu_{n'}
-\frac{d}{d\tau} \Big(
\frac{\partial A^\nu}{\partial \dot{x}_{n'}^\mu}
\Big) 
\nonumber \\
&- \frac{\partial A_\mu}{\partial x_{n';\nu}}
+ \frac{\partial A^\nu}{\partial x_{n'}^\mu}
\Big]
+D_{\;\xi_{n,i}}^\mu \Big[
\frac{d}{d\tau} \Big(
\frac{\partial A^\nu}{\partial \dot{x}_{n}^\mu}
\Big) 
+ \frac{\partial A_\mu}{\partial x_{n';\nu}}
- \frac{\partial A^\nu}{\partial x_{n}^\mu} 
  \Big]
\Big\} \dot{x}_{n';\nu}
\nonumber \\
&+ \Big\{
 D_{\;\xi_{n',i}}^\mu \Big[ 
\Big(m_{n'}(0) + \frac{\partial m_{n'}}{\partial x^\alpha_{n'}} {x}_{n'}^\alpha
+ \frac{\partial m_{n'}}{\partial \dot{x}^\alpha_{n'}}  \dot{x}_{n'}^\alpha \Big) g^\nu_\mu 
- \frac{\partial A^\nu}{\partial x_{n'}^\mu}
  \Big]
\nonumber \\
& + D_{\;\xi_{n,i}}^\mu \Big[ 
\frac{\partial A^\nu}{\partial x_{n}^\mu}
+\frac{\partial A_\mu}{\partial x_{n';\nu}}
\Big]
\Big\} \ddot{x}_{n';\nu}
+ D_{\;\xi_{n',i}}^\mu \Big(\square_{n'}^{\dot{\alpha}} m_{n'} \ddot{x}_{n';\alpha}\Big)\dot{x}_{n';\mu}.
\label{RotConstraintEqApprx}
\end{align}
The final extended Lagrangians have the form:
\begin{align}
L_T &=  \sum_{n}  \frac{1}{2}\big(m_n(0) + \frac{\partial m_n}{\partial x^\mu_n}  x_n^\mu 
+ \frac{\partial m_n}{\partial \dot{x}^\mu_n}  \dot{x}_n^\mu \big) \dot{x}^\nu_{n} \dot{x}_{n;\nu} -\big(A^\nu(0) + \frac{\partial A^\nu}{\partial x^\mu_n} x_n^\mu 
+ \frac{\partial A^\nu}{\partial \dot{x}^\mu_n} \dot{x}_n^\mu \big) \dot{x}_{n;\nu} 
\nonumber \\ 
&+ \omega_n ( x^\nu_n x_{\nu;n} -R^2_n) + \epsilon_n \partial_{n;\nu} A^\nu
+  \eta^\mu_n \Big[ \frac{\partial m_n}{\partial \dot{x}^\nu_n} \ddot{x}^\nu_n \dot{x}_{n;\mu} + 
\Big(m_n(0) + \frac{\partial m_n}{\partial x^\mu_n}  x_n^\mu 
+ \frac{\partial m_n}{\partial \dot{x}^\mu_n}  \dot{x}_n^\mu \Big)\ddot{x}_{n;\mu} \Big]
\nonumber \\ 
&+ \sum_{n'\neq n} \lambda^\mu_{n} \Big[  (m_{n'}(0) 
+ \frac{\partial m_{n'}}{\partial x^\alpha_{n'}} x_{n'}^\alpha
+ \frac{\partial m_{n'}}{\partial \dot{x}^\alpha_{n'}}  \dot{x}_{n'}^\alpha) \ddot{x}_{{n'};\mu} 
+ (\square_{n'}^\nu m_{n'} \dot{x}_{{n'};\nu}) \dot{x}_{{n'};\mu} 
- \frac{\partial A^\nu}{\partial \dot{x}_{n'}^\mu}\ddot{x}_{{n'};\nu} 
\nonumber \\ 
&- \frac{d}{d\tau}\Big(\frac{\partial A^\nu}{\partial \dot{x}_{n'}^\mu} \Big) \dot{x}_{{n'};\nu}
- \frac{1}{2}\frac{\partial m_{n'}}{\partial x_{n'}^\mu}\dot{x}^\nu_{n'} \dot{x}_{{n'};\nu} 
+ \frac{\partial A^\nu}{\partial x_{n'}^\mu}\dot{x}_{n';\nu}
+ \frac{d}{d\tau} \Big(\frac{\partial A^\nu}{\partial \dot{x}_{n}^\mu}\Big) \dot{x}_{n';\nu} 
\nonumber \\ 
&+ \frac{\partial A^\nu}{\partial \dot{x}_{n}^\mu} \ddot{x}_{n';\nu}  
- \frac{\partial A^\nu}{\partial x_{n}^\mu}\dot{x}_{n';\nu}
+ \square_{n'}^{\dot{\mu}} m_{n'} \ddot{x}_{n';\mu}  \dot{x}_{n';\nu}
\Big]   \label{extTransLagrangianAppx1}
\end{align}
and 
\begin{align}
L_R &=  \sum_{n} \frac{1}{2}\big(m_n(0) + \frac{\partial m_n}{\partial x^\mu_n}  x_n^\mu
+ \frac{\partial m_n}{\partial \dot{x}^\mu_n}  \dot{x}_n^\mu \big) \dot{x}^\nu_{n} \dot{x}_{n;\nu} 
-(A^\nu(0) + \frac{\partial A^\nu}{\partial x^\mu_n} x_n^\mu 
+ \frac{\partial A^\nu}{\partial \dot{x}^\mu_n} \dot{x}_n^\mu ) \dot{x}_{n;\nu} 
\nonumber \\ 
&+ \omega_n ( x^\nu_n x_{\nu;n} -R^2_n) + \epsilon_n \partial_{n;\nu} A^\nu
+  \eta^\mu_n \Big[ \frac{\partial m_n}{\partial \dot{x}^\nu_n} \ddot{x}^\nu_n \dot{x}_{n;\mu} + 
\Big(m_n(0) + \frac{\partial m_n}{\partial x^\mu_n}  x_n^\mu 
+ \frac{\partial m_n}{\partial \dot{x}^\mu_n}  \dot{x}_n^\mu \Big)\ddot{x}_{n;\mu} \Big]
\nonumber \\ 
&+\sum_{i,n'\neq n} \beta_{i_n} \Big\{  D_{\;\xi_{n',i}}^\mu \Big[ \big(m_{n'}(0) 
+ \frac{\partial m_{n'}}{\partial x^\alpha_{n'}} x_{n'}^\alpha 
+ \frac{\partial m_{n'}}{\partial \dot{x}^\alpha_{n'}}  \dot{x}_{n'}^\alpha \big) \ddot{x}_{{n'};\mu} + \big(\square_{n'}^\nu m_{n'} \dot{x}_{{n'};\nu}\big) \dot{x}_{{n'};\mu} 
\nonumber \\
&- \frac{\partial A^\nu}{\partial \dot{x}_{n'}^\mu}\ddot{x}_{{n'};\nu} 
- \frac{d}{d\tau}\Big(\frac{\partial A^\nu}{\partial \dot{x}_{n'}^\mu} \Big) \dot{x}_{{n'};\nu}
- \frac{\partial A_\mu}{\partial x_{n';\nu}}\dot{x}_{n';\nu} 
- \frac{\partial A_\mu}{\partial \dot{x}_{n';\nu}}\ddot{x}_{n';\nu}  
\nonumber \\
&- \frac{1}{2}\frac{\partial m_{n'}}{\partial x_{n'}^\mu}\dot{x}^\nu_{n'} \dot{x}_{{n'};\nu} 
+ \frac{\partial A^\nu}{\partial x_{n'}^\mu}\dot{x}_{n';\nu} \Big] 
+ D_{\;\xi_{n,i}}^\mu \Big[ \frac{d}{d\tau} \Big(\frac{\partial A^\nu}{\partial \dot{x}_{n}^\mu}\Big) \dot{x}_{n';\nu} 
+ \frac{\partial A^\nu}{\partial \dot{x}_{n}^\mu} \ddot{x}_{n';\nu}
\nonumber \\
&- \frac{\partial A^\nu}{\partial x_{n}^\mu}\dot{x}_{n';\nu} 
+ \frac{\partial A_\mu}{\partial x_{n';\nu}}\dot{x}_{n';\nu} 
+ \frac{\partial A_\mu}{\partial \dot{x}_{n';\nu}}\ddot{x}_{n';\nu}  \Big]
+ D_{\;\xi_{n',i}}^\mu \Big(\square_{n'}^{\dot{\alpha}} m_{n'} \ddot{x}_{n';\alpha}\Big)\dot{x}_{n';\mu}
\Big\} ,
 \label{extRotLagrangianAppx1}
\end{align}
%\end{widetext}
Since both Lagrangians are needed for describing the physical system, we can write them as a single two-components Lagrangian as
\begin{equation}
L_{sys}^* \equiv 
\classoperator{L_T}{ L_R}= 
\classoperator{L_{Rel} + \sum_{n} \lambda^\nu_n \Phi_{\nu_n}}
{L_{Rel} + \sum_{i,n} \beta_{i_n} \Psi_{i_n}},\label{extLagran}
\end{equation}
where we introduce the 7-$n$ independent constants $\lambda^\nu_n$ and $\beta_{i_n}$. Note that, while $\lambda^\nu_n$ constants need to be included in the covariant form, $\beta_{i_n}$ constant is invariant under Lorentz transformation. We use the matrix notation for expressing the general Lagrangian.

Thereby, the solutions that extreme Hamilton's least action principle for $n$-VMVF systems,  using the extended Lagrangian expressions eq. \ref{lagrangetranslationExt} and \ref{lagrangeRotationExt1}, can be written as
\begin{align}
&\mathcal{L}_{n,\mu,\xi_i}L_{sys}^*
= 
\begin{bmatrix}
\mathcal{L}_{n,\mu} & 0 \\ \\ 0 & \mathcal{L}_{n,\xi_i}
\end{bmatrix}
\classoperator
{L_T}
{L_R}
%\nonumber \\
%&=
%\classoperator{
%\mathcal{L}_{n,\mu} L_T}{
%\mathcal{L}_{n,\xi_i} L_R} =0
\nonumber \\
&=\classoperator{
\Big[ - \frac{d^2}{d\tau^2}\frac{\partial}{\partial \ddot{x}_n^\mu} 
+ \frac{d}{d\tau}\frac{\partial}{\partial \dot{x}_n^\mu} 
- \frac{\partial}{\partial x_n^\mu}\Big] L_T}
{
D_{\;\xi_{n,i}}^\mu \Big[ -  
\frac{d^2}{d\tau^2}\Big(  \square_{n,\ddot{\mu}} \Big) + 
\frac{d}{d\tau}\Big(  \square_{n,\dot{\mu}} \Big) - 
\square_{n,\mu} \Big] L_R} \label{extEulerLagranEq}
\end{align}
and
\begin{align}
&\Omega_{n,\mu,\xi_i} \equiv 
\classoperator
{\Phi_{\nu_n}}
{\Psi_{i_n}} 
= 0. \label{extEulerLagranConst}
\end{align}

Under an external field with a defined form and depending on the position of particles, the extended Lagrangian equation should have the form:
\begin{align}
\mathcal{L}_{n,\mu,\xi_i}L \equiv 
\classoperator{
\mathcal{L}_{n,\mu} (L_T - V(x^\nu))}
{\mathcal{L}_{n,\xi_i} (L_R - V(\xi_i))} =0
\nonumber \\
\Omega_{n,\mu,\xi_i} \equiv 
\classoperator
{\Phi_{\nu_n}}{ 
\Psi_{i_n}} = 0
\label{extEulerLagranEqV}
\end{align}

\section{Conclusions}
We have shown a new proposal for a classical theory for $n$ particle systems with variable masses connected by a field with no predefined form ($n$-VMVF systems). The proposition shows that it is possible to consider the masses and the field as unknown functions on the positions and velocities of the particles, and found them a solution to the problem using only the first principles.

The solution involves the resolution of not one but two set of Lagrange equations in the Lorentzian and the angular coordinates respectively. The answer shows that the problem of treating the mass and field as unknown functions is inherently relativistic. The four-dimensional space-time, and with it, the theory of Special Relativity, is naturally included in the problem because of the need of expressing the position of the particle as a function of angular coordinates. Different from the reason for adding the Theory of Special Relativity on other classical theories, in this problem the $3-D$ space of the angular coordinates is shown to be the stereographic projection of the $4-D$ sphere defined by the Lorentz condition in the $4-D$ space-time.

The obtained solution indicates the need for constructing a new Hamilton theory. The new classical theory should be composed of two set of constrained second order Hamilton's equations from where can be obtained the canonical transformations. The new Hamilton Theory should guide the definition of the quantum operator for the construction of a new Quantum theory which includes the mass and also the field as variable quantities to be determined in the final solution. 

The wanted solution to the problem is complex and extensive, but it can be seen as the starting point for more solvable approaches. Also, because of considering the masses and field as unknown quantities, the difference between $n$-VMVF systems lays only on the number of particles. That means that once we solve the classical problem for a particle number of particles, we will obtain a universal Lagrangian for that number of particles that can be used for solve any problem with that number of particles.

The theory permits the existence of particles with zero mass, which, in the quantum approach at the ground state, might be related to the vacuum. The incorporation of such particles would allow the study of the interaction of massive particles with the vacuum. In this new framework, we might re-examine some concepts and phenomena like spin, which can be related to the quantum angular momentum of the $2$-VMVF system composed of one massive particle and one vacuum particle. An expected result of the quantification of the mass and the field will be the obtainment of the masses of all known elementary particles and all fundamental interactions.

\section{acknowledgments}
I would like to express my deep gratitude to my mentor and personal friend Professor Dr. Fernando Guzm\'an Mart\'inez, for their guidance, encouragement, and critiques. I like to acknowledge professor Dr. A. Deppman for his teaching, advice and comments on this work. I also recognize the support from Dr. Yoelvis Orozco, Dr. Juan A. Garc\'ia, Dr. Yansel Guerrero and Dr. Rodrigo Gester.
\bibliography{References}

\end{document}